\documentclass[iop]{emulateapj}

\usepackage{graphicx}
\usepackage{color}
\usepackage{amsmath,amssymb}
\usepackage{type1cm}
\usepackage{ulem}
\usepackage{mathrsfs}

\slugcomment{draft version \today, }

\shorttitle{Photospheric Emission From Stratified Jets}                                                      
\shortauthors{Ito et al.}

\begin{document}

\title{Spectral and Polarization Properties of  Photospheric Emission From Stratified Jets}

\author{Hirotaka Ito\altaffilmark{1}, Shigehiro Nagataki\altaffilmark{1},  Jin Matsumoto\altaffilmark{1}, Shiu-Hang Lee\altaffilmark{2},   Alexey Tolstov\altaffilmark{3,4}, Jirong Mao\altaffilmark{1,5}, Maria Dainotti\altaffilmark{1} and Akira Mizuta\altaffilmark{6}}

\altaffiltext{1}{Astrophysical Big Bang Laboratory, RIKEN, Saitama 351-0198, Japan}
\altaffiltext{2}{Institute of Space and Astronautical Science, Japan Aerospace Exploration Agency, 3-1-1 Yoshinodai, Chuo-ku, Sagamihara, Kanagawa 252-5210, Japan}
\altaffiltext{3}{Kavli IPMU (WPI), University of Tokyo, Chiba 277-8583, Japan}
\altaffiltext{4}{Institute for Theoretical and Experimental Physics (ITEP), 117218, Moscow, Russia}
\altaffiltext{5}{Department of Physics, Kyushu University, Fukuoka 812-8581, Japan}
\altaffiltext{6}{Computational Astrophysics Laboratory, RIKEN,Saitama 351-0198, Japan}
\email{hirotaka.ito@riken.jp}

\begin{abstract}
We explore  the spectral and polarization
 properties of 
photospheric emissions from stratified jets
in which multiple components, separated by a
sharp velocity shear regions, are distributed in lateral direction.
Propagation of thermal photons injected at high optical depth region
are calculated until they escape from the photosphere.
It is found that 
presence of the
lateral structure  within the jet 
leads to non-thermal feature of the spectra
and significant polarization 
signal in the resulting emission.
The deviation from  thermal spectra
as well as the polarization degree
tends to be enhanced as the
velocity gradient in the shear region increases. 
In particular, we show that
emissions from multi-component jet
 can reproduce the typical observed spectra
 of gamma-ray bursts (GRBs)
 irrespective to the 
position of the observer
when a velocity shear region is closely 
spaced in various lateral ($\theta$) positions.
The degree of polarization
associated in the emission 
is significant ($>{\rm few} \%$) at wide range of 
observer angles and can be higher than  $30\%$.
\end{abstract}


\section{INTRODUCTION}

Gamma-ray Bursts (GRBs) are the most luminous phenomena in the 
Universe.
They are  transient, intense flashes of gamma-rays
occurring at cosmological distances.  One of their peculiar features is 
the rapid variability in their prompt emission light curves.
The observed spectrum is highly non-thermal and it is often
described by the empirical
Band function \citep{BMF93}
that has a smoothly  jointed broken power-law shape.
The typical break (peak) energy
is observed around $\sim0.1-1~{\rm MeV}$,
while the typical photon indices of the 
low and high energy spectrum are found at
$\alpha_{\rm ph} \sim -1$ and $\beta_{\rm ph} \sim -2.5$, respectively
\citep{PBM00,KPB06,KGP08, NGG11, GBP12, GPM13}.

It is widely accepted that the prompt emission is 
originated in an ultra-relativistic jet.
However,
despite the extensive studies in the past decades,
exactly how the gamma-rays are produced within the jet
remains unclear.
Optically thin synchrotron emission originating
from internal shocks \citep{RM94,SP97} has been 
considered as a standard model for many years.
In this model, highly non-thermal features 
showing a broken power-law 
and the observed rapid time variability
can be  achieved naturally.
On the other hand, however, it is known 
that this model faces several difficulties.
Since the internal shocks can only convert
the kinetic energy associated with the relative motion within the jet
into the gamma-rays,
it suffers from poor radiation efficiency
\citep{KPS97, LGC99, GSW01, KMY04}.
Additional difficulties are also found in the spectra.
A natural mechanism to realize 
clustering of peak energy at $\sim 0.1-1~{\rm MeV}$ is uncertain.
Moreover, the synchrotron model predicts the low energy photon index at
 $\alpha_{\rm ph} \sim -3/2$ for the electrons in fast cooling regime
which  conflicts  with  the observed typical value ($\alpha_{\rm ph}\sim -1$).
Furthermore,
 non-negligible fraction of the observed bursts
shows low energy slope harder than the death line $\alpha_{\rm ph}=-2/3$
which cannot be produced by usual synchrotron emission \citep{CLS97, PBM98, GCG03}.

Due to these difficulties,
photospheric emission model 
is considered as
one of the most promising alternative scenario
for the prompt emission mechanism 
\citep[e.g.,][]{T94, EL00,
MR00, RM05, LMB09,  PR11, MNA11, NIK11, RSV11, XNH12, BSV13, LPR13a, LMM13}.
The photospheric emission is an
inherent feature of the original fireball model
in which
the internally trapped photons that accelerate 
outflow to ultra-relativistic velocity 
are eventually released at the photosphere
\citep{G86,P86}.
Unlike in the internal shock model,
high radiation efficiency 
as well as clustering of the peak energies 
can be achieved quite naturally.
%
Recent detections of quasi-thermal component in the
observed GRB spectra provide a further support
that the prompt gamma-rays, at least in part, are originated
from the photosphere \citep[e.g.,][]{AAA09, RAZ10, RPN11, GCB11, PZR12, GDH13, GPG13}.

On the other hand, 
since thermal photons are expected in a
simple photospheric emission model, it is difficult to
reproduce 
the broad non-thermal shape of the observed spectra.
Hence, in order to explain the overall spectrum
with the photospheric emission alone,
an additional mechanism which leads to the broadening in the spectra
is required.
Several authors
claim an efficient dissipation
 around the photosphere
 can provide such an effect
\citep[e.g.,][]{PMR05, PMR06, G06, GS07, G08, IMT07, LB10, B10, VBP11, AM13}. 
The electrons (and positrons) that are heated
via dissipative processes such as shocks, 
magnetic reconnection 
and proton-neutron collisions 
can give rise to the non-thermal spectrum close to the observations.
However, it seems
quite questionable
whether
dissipative processes can operate efficiently  to
deposit  the copious amount of energy
in relativistic electrons (and positrons).
%

Alternatively, 
recent studies have shown that
broadening of spectrum can also be originated
by the effect of geometrical structure of the jet.
By considering a
gradually decaying profile in the
 lateral distribution  of the bulk Lorentz factor
 of the jet, \citet{LPR13a} found that
the typical low energy photon index ($\alpha_{\rm ph} \sim -1$)
can be reproduced by the photospheric emissions.
Similar to their approach,
but with a larger gradient (velocity shear) in the lateral profile,
 \citet{INO13} (hereafter, Paper I) have shown that,
by imposing a sharp gradient on the lateral distribution,
fraction of photons that cross the velocity shear region
multiple times can gain energy via Fermi-like acceleration mechanism.
It is demonstrated that the 
accelerated photons can give rise to non-thermal tail above the peak energy
which  reproduces the typical high energy photon index
 ($\beta_{\rm ph} \sim -2.5$) 
of the observed spectra.
However,
how the structure of the jet 
is naturally regulated to such a geometry, which can reproduce the 
observations, remains unclear.
%

In addition to the spectral features,
polarization measurement may be
useful to give a further constraint on the prompt emission mechanism.
The first detection of linear polarization in the prompt 
emission was reported by RHESSI 
from GRB 021206 \citep{CB03}.
However, independent groups did not confirm the polarization 
signals using same data \citep{RF04, WHA04}.
Similarly,
INTEGRAL-SPI and -IBIS data showed detection of polarization 
from  GRB 041219,
but  the results of SPI and IBIS appear inconsistent
\citep{KBK07, MCD07, GLL09}.
The instrumental systematics are the main obstacles to obtain
a convincing result.
%
Recent observation by
GAP instrument on board  IKAROS realized
a detection with quite low systematic uncertainty \citep{YMGG11}.
High degree of linear polarization 
in 
the prompt emission of GRB 100826A ($27\pm 11 \%$),
GRB 110301A ($70 \pm 22 \%$), and GRB 110721A ($84^{+16}_{-28} \%$)
were reported in the observations \citep[][]{YMG11, YMG12}.
It is noted, however, that these results still suffer from low statistics,
and $\sim 0\%$ polarization degree cannot be ruled out at
$\sim 3 \sigma$ confidence level.
%
Future polarimeter missions, such as TSUBAME \citep{YHK12}
and POLAR \citep{OPOL11},
are awaited to provide more accurate data.

In order to use the polarization measurement as a probe
for the prompt emission mechanism, 
it is essential to evaluate the
polarization signal associated in each emission models.
The polarization properties of 
optically thin models such as
synchrotron and
inverse Compton
emissions
have been
extensively studied by many authors
\citep[e.g.,][]{SD95, LPB03, G03,   NPW03, W03, EL03, LE04, LRG04, TSZ09, L06, ZY11, MW13}.
On the other hand,
only  few studies explored the 
polarization signals associated with 
photospheric emissions. 
First detailed study
on this issue was 
carried out by \citet{B11}.
He solved a transfer of photons 
within a steady relativistically expanding opaque outflow,
under the assumption of spherical symmetry.
Contrary to naive expectation,
it is found that significant
anisotropy develops in the photon distribution
around the  photosphere not only in the laboratory frame 
but also in the fluid comoving frame.
Due to the anisotropy,
photons released at the photosphere
can achieve high level of polarization through last scatterings.
However, although the photons released at a local emitting regions
can be strongly polarized, superposition of each emission component 
vanishes the net polarization in the observed emissions 
for a spherically symmetric outflow.
Therefore,
to produce a
non-negligible polarization signal in the observed emission,
the emitting region must show a
break of rotational symmetry  around the line of sight of the observer
within an angle $\sim \Gamma^{-1}$, where
$\Gamma$ is the bulk Lorentz factor of the outflow.

This is indeed shown by \citet{LPR14}
who studied the polarization properties of the  
photospheric emission from a jet having lateral structure.
The considered geometry 
is identical to those considered in their 
previous study \citep{LPR13a} which focused on the spectral features.
The bulk Lorentz factor of the jet
near the center is approximately constant 
up to a certain angular width 
and an approximately power-law decaying profile is imposed
at a larger angle  $\Gamma \propto \theta^{-p}$. 
By solving the photon transfer,
they found that
 a significant polarization signal
can be accompanied in  photospheric emissions originating from
a jet with such a geometry.
Particularly, they showed that
a narrow jet that has a steep Lorentz factor
gradient at the outer region 
can produce a quite high polarization degree
up to $\sim 40 \%$.
%
It is noted, however,
that  high polarization degrees $\gtrsim 10\%$ can be 
detected only by an observer that has
 line of sight, hereafter called LOS,
aligned in the outer regions.
The emissions viewed by such an off-axis observer
are much dimmer than
those observed by an on-axis observer.
%
Moreover, while they showed that a 
highly non-thermal  broken-power law shape
of the spectra is an inherent feature of 
the emissions,
the spectral slopes 
have a strong dependence on the observer position
particularly at high energies
and may deviate largely from the 
typically observed ones at a large fraction of observers.


As discussed in Paper I,
the strong dependence of the spectral slopes
on the observers can be reduced when a velocity shear
regions are present at various lateral position 
of the jet.
For example,
if a velocity shear region is distributed 
within the entire jet, 
non-thermal 
photons 
originated at the 
velocity shear regions due to photon acceleration
can be prominent for all observers.
Indeed, such a rich internal structure within the jet is inferred
in a recent numerical simulations \citep{MM13a, MM13b}.
In these simulations, they explored the
evolution of the transverse structure of the relativistic jet
during its propagation in a dense medium.
Their results indicated that
hydrodynamical instabilities such as Rayleigh-Taylor
and Richtmyer-Meshkov instabilities
produce small scale filamentary structures
that have sharp interface to distribute in entire jet regions.

Motivated by these backgrounds, 
we explore the photospheric emissions 
from a stratified jet 
that has velocity shear regions at various angular ($\theta$) positions
in the present study.
Main purpose of our study is to 
further investigate the effect of jet geometry 
on the spectral and polarization properties.
%
Analogously to Paper I,
here it is assumed that components with uniform fluid
properties are separated by a sharp boundary transition layers.
%
%
We show, in particular,
that, when a velocity shear region is closely 
spaced within an angular scale $\sim 2\Gamma^{-1}$,
typical Band spectra
can be reproduced irrespective to
the observer position.
We also show that
a high polarization degrees
$\gtrsim 10\%$
can be detected
not only by the observers viewing dim emissions, 
but also those viewing brightest emissions.




The paper is organized as follows. 
In \S\ref{model}, we describe our model and numerical procedures
used in our calculations.
We present the main results in \S\ref{result}.
Discussions are given in
 \S\ref{discussions}.
The summary of our main findings is given in \S\ref{summary}.

\section{MODEL AND METHODS}
\label{model}

In the present study,
we evaluate the photospheric emissions from
an ultra-relativistic jet
with a half-opening angle $\theta_{\rm j}$
in
which a stratified  structure is present in the
lateral ($\theta$) direction.
%
We consider two types for the stratification: 
(I) two-component (spine-sheath)
 jet in which
fast spine jet is embedded in a slower sheath outflow 
and 
(II) multi-component jet which is composed of multiple outflow
layer of finite lateral width.
The schematic picture of the two models are shown in 
Figs.~\ref{modeltwo} and \ref{modelmul}.

In both models,
we assume a sharp transition layer
between each component that has lateral width $d\theta_{\rm B}$.
As for the two-component jet model, the spine region 
defined as a region of conical outflow with a
half-opening angle $\theta_0- d\theta_{\rm B}/2$,
while
the sheath is a region that
surrounds the spine 
and has an angular extension of
 $\theta_0 + d\theta_{\rm B}/2 \leq \theta \leq \theta_{\rm j}$.
In the multi-component jet model, 
two components  
having fixed  widths of 
$d\theta_0-d\theta_{\rm B}$ (Component 0, hereafter C0) and 
$d\theta_1-d\theta_{\rm B}$ (Component 1, hereafter C1) alternately appear
in the transverse direction.
While the first component located at the 
center is a conical outflow
with a half-opening angle $(d\theta_0-d\theta_{\rm B})/2$,
other components
have sheath structures 
 which have the same central axis.
The repeated pattern of this transverse structure continues
until the total angular extension reaches the half opening angle of the
jet, $\theta_{\rm j}$.
As  described below, the properties  of the 
spine (sheath) in the two-component jet model
are determined in the same manner as
C0 (C1) in the multi-component jet model.
%
Hereafter, 
the quantities corresponding to the spine
 (C0) and sheath (C1)
regions are denoted by subscripts $0$ and $1$, respectively.
The quantities without the subscript refer to all regions including the
boundary transition layers.

It is noted that the two-component jet model
is introduced to clarify the effects of 
sharp velocity gradients on the resulting spectra and the polarization.
Main difference
between the present paper and Paper I
 is that here we include
and quantify the polarization signal.
On the other hand,  
multi-component jet model is introduced 
in order to explicitly
show that the 
existence of sharp velocity gradient regions 
within an angular scale $\sim 2\Gamma^{-1}$
is essential to reproduce the typical observed spectra of GRBs
 and
to quantify the polarizations associated with these emissions.
Although the assumed geometry of the employed models
is somewhat artificial,
it is stressed that 
similar results are expected
if sharp velocity shear regions are present in the transverse structure 
of the jet and
are closely distributed 
within a small angular scale $\sim 2\Gamma^{-1}$.
Such a rich internal structure is indeed inferred from
the recent numerical simulations \citep[e.g.,][]{MM13a, MM13b}.
We will mention on this issue later in \S\ref{Multi} 
and \S\ref{stdis}.

\subsection{Fluid Properties of Stratified Jet}
\label{PTSJ}

We consider a steady
radially expanding axisymmetric outflow, and
the radial profile
of the fluid properties
 are described by the standard adiabatic fireball
model \citep[e.g.,][and also see Paper I for a brief review]{P04,M06}
which can be determined uniquely by the three independent parameters:
the initial radius, $r_{\rm i}$,
the kinetic luminosity, $L$, and
the dimensionless entropy (or, equivalently the terminal Lorentz factor)
$\eta \equiv L/\dot{M}c^2$, where $\dot{M}$ and $c$ are
 the mass outflow rate and the speed of light, respectively.
In the present study, we only consider the case in which
the photosphere, $r_{\rm ph}$, the radius where
the fireball becomes optically thin (see Eq.~(\ref{rph})),
is located above the saturation radius, $r_{\rm s}=\eta r_{\rm i}$,
the radius where the bulk acceleration of the fireball ceases.
Hence, the three parameters in all regions satisfy
$\eta < \eta_c= (\sigma_{\rm T} L / 8\pi r_{\rm i} m_p c^3)^{1/4}$, 
where $\sigma_{\rm T}$ and $m_p$ are the
Thomson cross section and proton rest mass, respectively.

Given the three parameters, the radial evolution 
of the bulk Lorentz factor 
and the electron number density is determine 
by
\begin{eqnarray}
\label{Gamma}
\Gamma(r) =   \left\{ \begin{array}{ll}
      \frac{r}{r_{\rm i}} &~~
         {\rm for}~~ r \leq r_{\rm s}  ,  \\
     \eta  &~~
         {\rm for}~~  r > r_{\rm s}  , \\
             \end{array} \right. 
\end{eqnarray}
and
\begin{eqnarray}
\label{ne}
n_e(r) = \frac{\dot{M}}{4 \pi r^2 m_p \Gamma \beta c}
 = \frac{L}{4 \pi r^2 m_p \eta \Gamma \beta c^3},
\end{eqnarray}
where $\beta$ is the velocity of the flow normalized by 
the speed of light. 
The
electron number density decreases with radius as
$n_e \propto r^{-3}$ below the saturation radius ($r\leq r_{\rm s}$) and
as $n_e \propto r^{-2}$ at larger radii ($r> r_{\rm s}$).
Given the electron number density and bulk Lorentz factor
of the flow,
the optical depth to Thomson scatterings for
the photons propagating in the {\it radial} direction
to reach infinity can be evaluated as
\begin{eqnarray}
\label{tau}
  \tau(r) &=&  \int^{\infty}_r \sigma_{\rm T} n_e(r') \Gamma(r')(1-\beta(r')) dr' \nonumber \\
&\simeq&
  \left\{ \begin{array}{ll}
    \frac{r_{\rm ph}}{r} \left[1+\frac{1}{3}
         \left\{ 
       \left(\frac{r_{\rm s}}{r} \right)^2
     -  \left(\frac{r}{r_{\rm s}} \right)^2 \right\} \right]  &~~
         {\rm for}~~  r \leq r_{\rm s}  , \\
      \frac{r_{\rm ph}}{r} &~~
         {\rm for}~~ r > r_{\rm s}  ,

  \\
             \end{array} \right. 
\end{eqnarray}
\begin{eqnarray}
\label{rph}
r_{\rm ph} = \frac{\sigma_{\rm T} L}{8 \pi \eta^3 m_p c^3},
\end{eqnarray}
where we have assumed $\Gamma \gg 1$ \citep{ANP91}.
Here, $r_{\rm ph}$ is 
the radius
where the optical depth becomes unity ($\tau=1$).

\subsubsection{Transverse structure}
\label{TrS}

Here, we describe how the transverse structure of the
jet  is determined.
The transverse structure 
plays an essential role 
on the spectral and polarization properties of the 
resulting emissions.

\begin{figure}[ht]
\begin{center} 
\includegraphics[width=9cm]{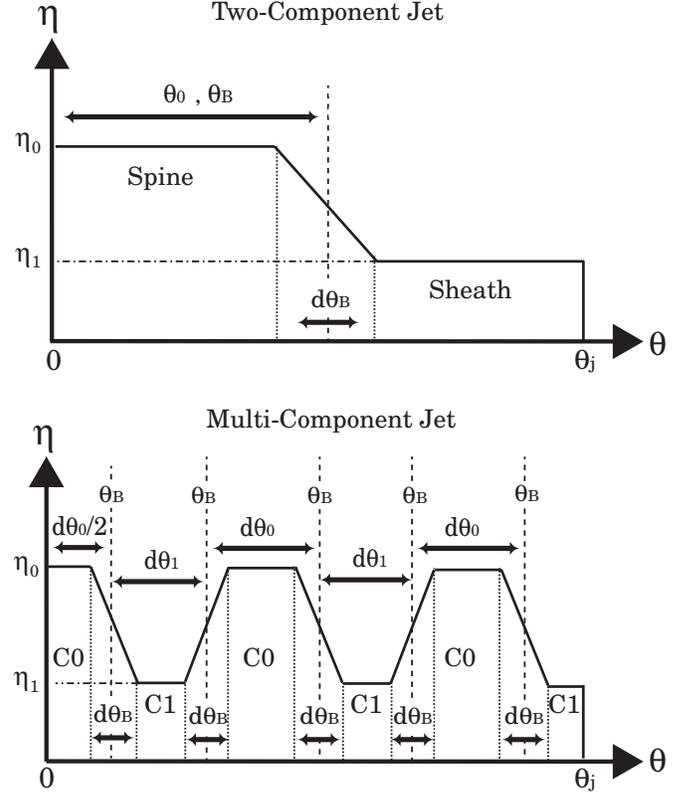}
\caption 
{The transverse ($\theta$) distribution of 
the dimensionless entropy (terminal Lorentz factor) $\eta$.
The top and bottom panels correspond to the
two-component and multi-component jet model, respectively.
} 
\label{modeleta}
\end{center}
\end{figure}

In all regions,
We impose the same value of the initial radius, $r_{\rm i}$. 
Different values are imposed for the 
dimensionless entropy of the
spine (C0), $\eta_0$ and 
sheath (C1) region,  $\eta_1$.
In all cases we assume $\eta_0 > \eta_1$.
As explained later,
 due to this difference, 
strong velocity shear region appears within the jet.
The transverse ($\theta$) distribution of the dimensionless entropy at the
boundary transition layer is determined by simply imposing 
a linear interpolation from the surrounding two regions
which can be written as
\begin{eqnarray}
\eta(\theta)=\frac{(\theta - \theta_{\rm B} + d\theta_{\rm B}/2)\eta_{\rm ii} + 
(\theta_{\rm B} + d\theta_{\rm B}/2 -\theta)\eta_{\rm i}}{d\theta_{\rm B}},
\end{eqnarray}
where $\theta_{\rm B}$ is the angle at midpoint of the transition layer.
The subscripts ${\rm i}$ and ${\rm ii}$  correspond to 
 $0$ ($1$) and $1$ ($0$), respectively,
when $\theta_{\rm B}-d\theta_{\rm B}/2$ and 
$\theta_{\rm B}+d\theta_{\rm B}/2$ are the angle at the 
edge of component with dimensionless entropy of 
$\eta_0$ ($\eta_1$) and $\eta_1$ ($\eta_0$), respectively.
Fig.~\ref{modeleta} shows
the overall transverse distribution of 
the dimensionless entropy (terminal Lorentz factor).

In determining the kinetic luminosities,
we assume that the mass outflow rate
is equal in all regions ($L / \eta = {\rm constant}$).
Hence, kinetic luminosity
of the spine (C0) is larger by a factor of $\eta_0/\eta_1$, than
that  of the sheath (C1).

Due to the difference in the imposed parameters,
the saturation radius of the spine (C0)
 $r_{\rm s0}=\eta_0 r_{\rm i}$ is located above 
that of the sheath (C1) $r_{\rm s1}=\eta_1 r_{\rm i}$.
Since $\eta(\theta)$ is determined by the linear interpolation
from the surrounded two regions,
the saturation radius at the layer is 
also a linear interpolation from $r_{\rm s0}$ and $r_{\rm s1}$.
As a result, 
the bulk Lorentz factor in all regions
has equal values 
($\Gamma(r) = r/r_{\rm i}$) up to $r=r_{\rm s1}$,
and, at a larger radii, ($r>r_{\rm s1}$), 
velocity shear begins to develop at the boundary layer
until $r=r_{\rm s0}$.
Thereafter, the velocity difference is fully developed
and all regions are in the coasting phase
with a terminal Lorentz factor
(see Fig.~\ref{modeleta}, for the resulting distribution).

Regarding the transverse distribution of the photosphere,
the photospheric radius 
in the spine (C0) region, $r_{\rm ph0}$,
is smaller than in the sheath (C1) region, $r_{\rm ph1}$,
by a factor of $(\eta_0/\eta_1)^2$ (see equation (\ref{rph})).
The photospheric radius in the boundary transition layer
also scales as 
$\propto (\eta_0/\eta(\theta))^2$ and
 continuously connects the surrounding two regions.

\begin{figure}[ht]
\begin{center} 
\includegraphics[width=8.5cm]{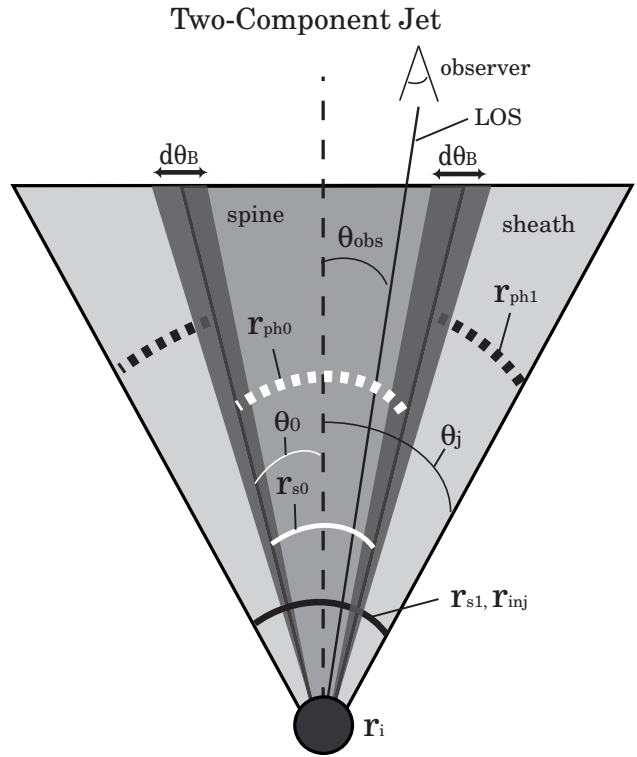}
\caption 
{Schematic picture of our two-component (spine-sheath) jet model.
 A fast spine jet ($\theta \leq \theta_0-d\theta_{\rm B}/2$)
 is embedded in a slower sheath outflow
 ($\theta_0 +  \theta_{\rm B} /2 \leq \theta \leq \theta_{\rm j}$).
 The spine and
 sheath start to accelerate at  radius $r_{\rm i}$.
 The acceleration continues up to $r_{\rm s0}$ and $r_{\rm s1}$
 in the spine and sheath region, respectively.
 Since the dimensionless entropy of the spine $\eta_0$ 
 is larger than that of the sheath $\eta_1$,  
 the saturation radius and the terminal Lorentz factor 
 of the spine ($r_{\rm s0} = \eta_0 r_{\rm i}$ and $\Gamma_0 = \eta_0$)
 are larger than those of the sheath 
 ($r_{\rm s1} = \eta_1 r_{\rm i}$ and $\Gamma_1 = \eta_1$).
 The photospheric radius of the spine $r_{\rm ph0}$ is smaller than that of the sheath $r_{\rm ph1}$, where the photospheric radius is 
 defined by Eq.~(\ref{rph}).
 There is a transition layer
 with an angular width $d\theta_{\rm B}$ 
 between the spine and
 sheath ($\theta_0 -  \theta_{\rm B} /2 \leq \theta \leq \theta_0 + \theta_{\rm B} /2$).
 The  dimensionless entropy and kinetic luminosity in this region 
 are determined by the interpolations from the two regions.
 In our calculation, thermal 
photons are injected at  the saturation radius of the sheath $r_{\rm inj} = r_{\rm s1}$, and their transfer is solved up to the radius at which
   the optical depth is much lower than unity.
} 
\label{modeltwo}
\end{center}
\end{figure}

\begin{figure}[ht]
\begin{center} 
\includegraphics[width=8.5cm]{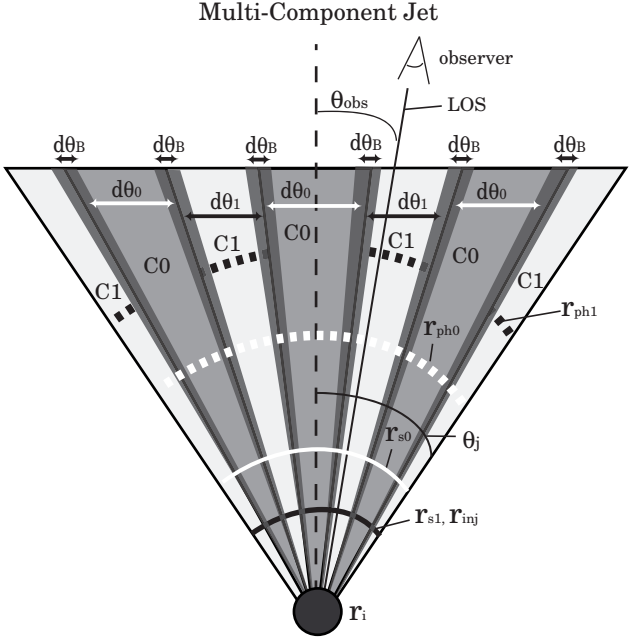}
\caption 
{Schematic picture of our multi-component jet model.
Two components  
having fixed  widths of 
$d\theta_0-d\theta_{\rm B}$ (C0) and 
$d\theta_1-d\theta_{\rm B}$ (C1) alternately appear
in the transverse direction   
within the jet with half opening angle $\theta_{\rm j}$
There are transition layers 
with an angular width $d\theta_{\rm B}$ between the two components. 
The radial profile of C0 and C1
as well as the transition layer 
are determined in the same way
as in the two-component
jet model.
} 
\label{modelmul}
\end{center}
\end{figure}

\subsection{Photon Transfer}
\label{MCcode}

Having determined the background fluid properties ($\Gamma$ and $n_e$),
we evaluate the resultant photospheric emission
by solving the propagation of photons.
%
The photon transfer is evaluated by performing a three-dimensional
test particle Monte-Carlo simulation.
In GRB jets, opacity of photons is strongly dominated by the scatterings 
with electrons \citep{PR11,B11}.
Therefore,
we neglect the absorption process and
only consider the scattering process by the electrons.
Furthermore, for simplicity,
 we do not take into account
the thermal motion of the electrons in evaluating the scattering.
Hence, the rest frame of the fluid is equivalent to that of the electrons.

The Monte-Carlo code used in the present study is basically the same
as the ones used in Paper I.
We directly track the path
of the individual photon packets that undergoes multiple-scatterings
with the electrons in the jets.
The main difference from the previous study is that 
we evaluate the polarization state of the photon packet,
and include its effect on the electron scattering.
Each photon packet carries a specified 
four momentum (or, equivalently the frequency, $\nu$,
 and the propagation direction ${\bf n}$).
In addition, 
the Stokes parameters $S=(I,Q, U, V)$ which determine
the polarization state
are also carried. 
The evolution of these quantities via electron
scattering is calculated self-consistently
by properly taking into account the
effect of the polarization state on the scattering process
(for detail, see 
Appendix).

The parameter $I$ corresponds to the intensity which we 
set to be equal to the total energy carried by  the packet.
The parameters $Q$ and $U$  characterize the linear
polarizations 
 with respect to arbitrary orthogonal $x-y$
axes in the plane of the polarization
 (plane that is perpendicular to the photon propagation direction)
and 
to a set of axes oriented at $45^{\circ}$ to the 
anti-clockwise direction
of the previous one, respectively.
The parameter $V$  characterizes the circular polarization.
However, $V$ is not relevant in our calculations
since we assume that the initially injected photons are unpolarized
($Q=U=V=0$) and that spin of the electrons are isotropically distributed.
Under these assumptions, scatterings only lead to 
the changes in the linear polarizations and $V=0$ is always conserved.
%

\begin{figure}[ht]
\begin{center} 
\includegraphics[width=8cm]{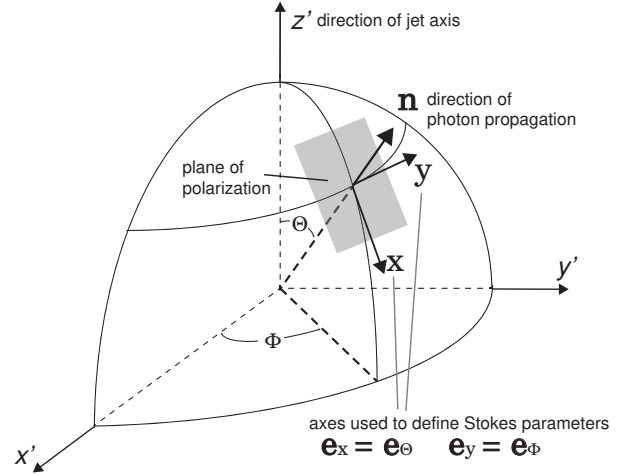}
\caption 
{The plane of polarization and 
 coordinate systems
 that are  used to define the Stokes parameters.
 The vector ${\bf n}$ shows the 
 direction of the photon propagation.
 The plane of polarization is defined as a plane 
 normal to ${\bf n}$.
 By introducing a Cartesian coordinate system
 that has its $z$-axis aligned to 
 the direction of the jet axis 
  (shown  by ${\it x}'$, ${\it y}'$ and ${\it z}'$),
 and the corresponding spherical coordinate system,
 the propagation direction of a photon can be defined by 
 two angles: $\Theta$, the angle between vector ${\bf n}$ 
 and the ${\it z}'$-axis, and 
 $\Phi$, the angle between the projected vector of ${\bf n}$
 in the ${\it x}'$-${\it y}'$ plane and the  ${\it x}'$-axis.
 The directions of the $x$- and $y$-axes that 
 are used to define the Stokes parameters are
 aligned to the direction of
 two basic  vectors of the spherical coordinate
 system ${\bf e}_{\Theta}$ and ${\bf e}_{\Phi}$, respectively.
} 
\label{axis}
\end{center}
\end{figure}

The $x$- and  $y$-axes used to
define the Stokes parameters are aligned in a plane
formed by a direction vector of photon ${\bf n}$ and
the jet axis
and to a direction perpendicular to the plane, respectively.
We illustrate the coordinates and polarization plane mentioned above in
 Fig.~\ref{axis}.
The photons propagating in the same direction have 
the same coordinate system, because even though the direction of the axes temporarily changes 
during the computation of the scattering effect on the 
polarization state (Appendix),  after the calculation,
 the Stokes parameters are always 
redefined in the above coordinate system.

Initially, the photons are injected within the jet 
at the surface of a fixed radius where the velocity
 shear begins to develop $r_{\rm inj} = r_{\rm s1}$.
For the cases considered in this study,
$r_{\rm inj}$ is always located far below the 
photosphere ($\tau(r_{\rm inj})\gg 1$). Therefore,
a tight coupling between the photons and matter is ensured.
For this reason, we can safely
assume that the photons
have an isotropic distribution
with energy distribution  given by a Planck distribution
 in the comoving frame and
these photons are unpolarized.

According to the fireball model, the radial evolution of the 
comoving temperature is given by
\begin{eqnarray}
\label{Tev}
T'(r) =   \left\{ \begin{array}{ll}
      \left( \frac{L}{4\pi r_{\rm i}^2 c a} \right)^{1/4}
      \left( \frac{r}{r_{\rm i}} \right)^{-1}
 &~~
         {\rm for}~~ r\leq r_{\rm s}  ,  \\
 \left( \frac{L}{4\pi r_{\rm i}^2 c a} \right)^{1/4}
      \left( \frac{r_{\rm s}}{r_{\rm i}} \right)^{-1}
      \left( \frac{r}{r_{\rm s}} \right)^{-2/3}
       &~~
         {\rm for}~~  r> r_{\rm s}  , \\
             \end{array} \right. 
\end{eqnarray}
where  $a$ is the radiation constant.
Hence,
we adopt the temperature at the corresponding radius
given by above equation  $T'_{\rm inj} = T'(r_{\rm inj})$
for the comoving temperature of the injected photons.
%
%
Due to this relativistic effect,
the radiation intensity  of the blackbody emission
 in a relativistically expanding flow 
is given by
\begin{eqnarray}
\label{Iin}
I_{\nu,{\rm inj}}(\nu) = {\cal D}(\Gamma_{\rm inj}, \theta_v)^3
 B_{\nu}(T'_{\rm inj}, \nu/{\cal D}(\Gamma_{\rm inj}, \theta_v)),
\end{eqnarray}
where
 $\Gamma_{\rm inj} = \Gamma(r_{\rm inj})$ 
is the bulk Lorentz factor of the flow
 at $r=r_{\rm inj}$ determined from equation (\ref{Gamma}).
Here, $B_{\nu}(T',\nu) =
2 h \nu^3 c^{-2} [{\rm exp}(h\nu)/k_{\rm B} T' - 1]^{-1}$ is the Planck function,
where
$h$ and
$k_{\rm B}$
are the Planck constant and the Boltzmann constant, respectively,
and
$D(\Gamma, \theta_v)=[\Gamma(1-\beta \cos{\theta_v})]^{-1}$ is the 
Doppler factor, where
$\theta_v$ is
 the angle between
the photon propagation direction and the fluid velocity direction
(radial direction).
In our calculation, 
the initial propagation direction and frequency of
the injected photons are drawn
from a source of photons given by the above equation.
Initially, 
the total energy of the packet $I$
is set to be equal among  all the photon packets.
As mentioned above,
the remaining Stokes parameters are set 
to be 0 ($Q=U=V=0$), since the photons far below the photosphere 
are expected to be unpolarized.

After the the photon packet is injected, 
we follow the track of the individual photons
using the Monte-Carlo technique.
Initially, 
the code determines 
 the distance for the photons to travel
before the scattering 
by drawing the corresponding optical depth  $\delta \tau$. 
The probability for the selected optical depth 
to be in the range of 
[$\delta \tau$, $\delta \tau + d\tau$] 
is given as ${\rm exp}(- \delta \tau) d\tau$.
Then, from the given optical depth $\delta \tau$,
the distance $l$ to the scattering event is
determined from the integration along the straight path
of photons which can be expressed as 
\begin{eqnarray}
\label{taucal}
\delta \tau = \int^l_0 n_e \Gamma(1-\beta \cos{\theta_v}) \sigma_{\rm sc} dl . 
\end{eqnarray}
Here, $\sigma_{\rm sc}$ is the total cross section for the electron 
scattering and is given as 
\begin{eqnarray}
\label{sccross}
\sigma_{\rm sc} =   \left\{ \begin{array}{ll}
      \sigma_{\rm T}
 &~~
         {\rm for}~~ h \nu_{\rm cmf} \leq  100~{\rm keV}  ,  \\ 
%
%
 \sigma_{\rm KN}
       &~~
         {\rm for}~~  h \nu_{\rm cmf}  >   100~{\rm keV}  ,  \\ 
             \end{array} \right.
\end{eqnarray}
in our code, where $\sigma_{\rm KN}$ is
the total cross section for  
Compton scattering, and
$\nu_{\rm cmf}$ 
is the frequency of the photon in electron (fluid)
comoving frame. 
The frequency $\nu_{\rm cmf}$
 is evaluated by performing a Lorentz transformation
 using local fluid velocity.

Given the distance $l$ from the above equation,
we update the position of the photons to the scattering position
by shifting them from the initial position
with the given distance
 along the  direction of photon propagation.
Note that, unlike the case of Eq.~(\ref{tau}),
the optical depth calculated by 
Eq.~(\ref{taucal}) is not limited to photons propagating  
in the radial direction.
The path of integration is along the straight path of photons
which can be in an arbitrary direction. 
For a given value of $\delta \tau$,
the distance $l$ strongly depends on the propagation direction of the photons
in the case of a relativistic flow ($\Gamma \gg 1$).
This is because
the mean free path of photons
 $l_{\rm mfp} = [n_e \Gamma (1-\beta \cos{\theta_v}) \sigma_{\rm sc}]^{-1}$ is
quite sensitive to the photon propagation direction,
since the factor $\Gamma (1-\beta \cos{\theta_v})$ varies
largely from $\sim (2\Gamma)^{-1}$ (for ${\rm cos}\theta_v = -1$) 
up to $\sim 2\Gamma$  (for $\cos{\theta_v} = 1$) depending on the 
value of $\theta_v$.
Hence, a  photon tends to travel a larger distance in 
the fluid velocity
(radial) direction, since the mean free path of the photon tends to be larger.  

Given the position for the scattering  from the above procedure, 
the four-momentum (the frequency and propagation direction)
and the Stokes parameters of a photon packet
after the scattering is determined
based on the differential cross section for 
electron scattering.
Here we  give the brief overview of our calculation
(for detail see Appendix).
In our code, 
the scattering process is evaluated in the
rest frame of the fluid. 
First,
the four-momentum of the photon before the scattering
is Lorentz transformed
from laboratory frame to the fluid rest frame.
For the photons that satisfy $h \nu_{\rm cmf} \leq   100~{\rm keV}$,
 the Klein-Nishina effect is neglected, while
 the full Klein-Nishina cross section is used at higher 
 energies ($h \nu_{\rm cmf} >   100~{\rm keV}$).
The effect of polarization on the differential cross sections
is self-consistently included in both regimes.
The scattering angle or equivalently
the propagation direction of the scattered photon
 in the fluid rest frame is drawn from the differential cross sections.
Once the scattering angle is determined,
we update the four-momentum and the Stokes parameters 
to that of the scattered photons.
 While the frequency  and total energy carried by a photon packet 
 ($\nu$ and $I$) are
 conserved after the scattering (elastic scattering)
 for $h \nu_{\rm cmf} \leq   100~{\rm keV}$,
 energy loss due to the recoil effect is taken 
 into account for 
$h \nu_{\rm cmf} >   100~{\rm keV}$.
 Also, the Stokes parameter of the scattered photon packet 
 is obtained from the incident one
 by properly taking into account the Klein-Nishina effect
 for $h \nu_{\rm cmf} >   100~{\rm keV}$,
 while the effect is neglected for 
 $h \nu_{\rm cmf} \leq   100~{\rm keV}$.
Then the four-momentum of the scattered photon
is transformed back into laboratory frame.

The above procedure is repeated
until all the injected photons
reach the outer boundary or inner boundary of the calculation.
The outer boundary in the radial 
direction is set at a radius $r_{\rm out}=2000r_{\rm ph0}$
where the photons can be safely
considered to have  escaped since the optical depth
in all region is
much smaller than unity 
($\tau(r_{\rm out}) \ll 1$).
The outer boundary in the
transverse ($\theta$) direction is set 
at the edge of the jet $\theta_{\rm out}=\theta_{\rm j}$.
As for the inner boundary, 
we adopt a radius slightly below the injection radius
 $r_{\rm in}=0.5r_{\rm inj}$.
For photons which have reached the outer boundaries,
we assume that they escape freely to $r=\infty$ without being
scattered or absorbed.
On the other hand, we assume that the photons are simply absorbed
in the inner boundary.
It is noted, however, that
the fraction of absorbed photons is negligible,
since  most of the photons in ultra-relativistic outflows 
are strongly collimated due to the relativistic beaming effect
 and essentially streamed outward \citep[e.g.,][]{PR11,B11}.

\subsection{Evaluation of the Observed Spectrum and Polarization}
\label{Eval}

After the calculations,
the spectra and the 
polarization of the emission are evaluated from the 
photon packets which have reached the outer boundaries.
Due to the relativistic beaming effect,
the emission depends strongly on
 the angle between the
direction to the observer and the jet axis, $\theta_{\rm obs}$
(see Figs.~\ref{modeltwo} and \ref{modelmul}). 
In the present study,
we evaluate the 
observed spectrum
and polarization 
by recording all escaped packets 
 that are propagating 
within a cone of half-opening angle $(4\eta_0)^{-1}$
around the LOS of a given observer.
The employed width of the cone  is small enough to regard that
the emission is uniform within the cone.
In evaluating the spectra,
for a given frequency bins,
we sum up the total energy carried by the photon
packets $I$ within the cone, and 
convert to the isotropic equivalent luminosity by 
multiplying it  with  a factor  $4\pi/d\Omega$,
where
 $d\Omega$ 
is the solid angle of the cone.
%
%
In the same way, 
all Stokes parameters ($I$, $Q$, $U$ and $V$) are
summed up in the given frequency bins and 
solid angles to evaluate
the corresponding Stokes parameters of the total emissions.
The polarization signal of the observed emission is 
evaluated from these parameters.
%
%

It is noted that the results of our calculation are insensitive to
the assumed position of the injection radius as long as
$r_{\rm inj} \leq r_{\rm s1}$ is satisfied, and
the observer angle, $\theta_{\rm obs}$,
(angle between the LOS and the jet axis) 
is  limited in the range
 $\theta_{\rm obs} \lesssim \theta_{\rm j}-\Gamma^{-1}$.
The former condition comes from the fact that,
at a radius far below the
photosphere ($\tau(r) \gg 1$),
the photon distribution 
does not
deviate from the isotropic Planck distribution and remains unpolarized, 
 if velocity shear is not present (see next section),
and its temperature evolution is well described by 
Eq.~(\ref{Tev}).  
The latter condition is required, since the 
photons near the jet edge escape
during their propagation.
Larger fraction of photons escape when smaller injection radius is assumed.
Hence, the 
resulting spectra and the polarization have 
dependence on the given injection radius.
However, it is noted that,
while the dependence is strong 
for an observer that have LOS located outside
the lateral jet boundary
($\theta_{\rm obs}>\theta_{\rm j}$),
 weak dependence is found at smaller observer angle,
 and quantitatively similar results are obtained 
when the injection radius is set far below the photosphere.
Therefore, in the present study, 
we focus on the observers that have LOS aligned within the 
jet region 
 $\theta_{\rm obs} \leq \theta_{\rm j}$
and do not consider the cases for  larger observer angles.

As mentioned above,
circular polarization is absent  and only linear polarization is found
in our calculation,
since the circular polarization parameter always satisfies $V=0$.
Also, due to the imposed axisymmetry and
the employed coordinate system for the Stokes parameters
(Fig.~\ref{axis}), 
the parameter $U$ vanishes after summing up \citep[see e.g.,][]{C60, B11}.
We have checked this in our calculation and indeed found 
that the parameter $U$ converges to $0$ 
as the number of photon packets increases.
The number of the packets employed in our calculation is 
sufficiently large to consider that this convergence is achieved.
Therefore, the non-zero Stokes parameters
are $I$ and $Q$. 
As a result, the polarization state of the  observed emissions are 
characterized by the two parameters and 
the degree of polarization given by $|Q|/I$.
When $Q$ is positive (negative), the electric vector of the polarized 
emission is aligned to $x(y)$-axis  shown in Fig.~\ref{axis}.
Hence, the positive $Q$
corresponds to the case for electric vector
aligned in the plane formed by photon propagation direction (LOS)
and jet axis,
while the  negative $Q$
corresponds to the case for that aligned to
 perpendicular to the plane. 

Finally, let us comment on the expected polarization
signature in the stratified jet.
As mentioned above, electron scattering of an unpolarized
photon packet results in a linearly polarized
outgoing photon packet.
In Thomson regime
(majority of scatterings in our calculation  occurs in this regime),
the degree of polarization
depends only on the scattering angle $\theta_{\rm sc}$,
and $100\%$ polarization is found when  $\theta_{\rm sc}=90^{\circ}$.
The electric vector of the scattered
photon is perpendicular to the plane in formed by  the incoming
and outgoing photon directions.
Hence,
emissions from
scattering dominated photosphere 
have a potential  to produce large linear polarization degree.

In order to produce
non-zero polarization degree
in the net polarization of the  emitted photons,
photon distribution in the comoving frame
must be anisotropic near the photosphere (last scattering surface).
This is because 
isotropically distributed 
photons do not have  preferential direction in the scattered photon 
field, and
all the polarization signal produced by
the scattering will vanish as a whole.
As shown by \citet{B11},
anisotropy is naturally produced
in a scattering dominated photosphere of
relativistically expanding fireball.
%
In a relativistically expanding outflow, 
mean free path of the photons has large dependence 
on the propagation direction.
The photons that propagate along the fluid velocity (radial direction)
have larger mean free path than those propagating in other direction.
This effect leads the photons to become concentrated in 
a radial direction.
On the other hand, 
scattering tends to reduce the anisotropy by 
re-randomizing the propagation direction of the photons.
When the photons are far below the photosphere, 
mean free path of the photons is small enough
to keep the photons to be isotropic
due to the latter effect.
However, when the photons reach near the photosphere 
the former effect can be significant to produce 
the  anisotropy in the comoving frame,
since the mean free path  becomes large.
As a result, substantial degree of polarization ($\gtrsim 10 \%$)
can be found in the photons released at the  last scattering surface.
The strongest polarization is found for 
the photons that propagate in the direction at
an angle $\sim 90^{\circ}$  respect to the 
radial direction in the comoving frame.
This is due to the fact, that the 
photons are mainly composed of 
population that have scattered in an angle close to $90^{\circ}$,
since the photon intensity is concentrated in the radial direction.
Hence, in an laboratory frame, 
the strongest polarization is found in the photons that 
are emitted (last scattered) at an angle $\sim \Gamma^{-1}$ 
respect to the radial direction.
The electric vector of the linearly polarized photons is
aligned to the direction perpendicular to
the plane formed by photon propagation direction and radial direction.

It is noted, however, that 
producing a polarized emissions at a local emitting region is
not a sufficient condition
for a  polarization signal  to be present
in  the detected emissions.
This is because 
the polarization  can vanish
when
contributions from the total emitting regions are summed up.
In an outflow that
is expanding in a radial direction with
a bulk Lorentz factor $\Gamma$,
most of the detectable emissions come from the 
regions within the cone of half opening angle $\sim \Gamma^{-1}$
around the LOS  of a given observer.
Hence, if the emission region is spatially axisymmetric
within the cone, the emitted polarized signatures 
cancel out as a whole.
Therefore, in addition, 
break of the axisymmetry in the emission region 
around the LOS within an angle $\sim \Gamma^{-1}$
is also a required condition for producing the detectable 
polarization  \citep[][]{B11, LPR14}. 

The stratified jet considered in the present
study satisfies  the above  two conditions, and 
therefore, substantial degree of polarization can 
be present in the observed emissions.


\section{RESULTS}
\label{result}
In this section, we show the obtained photon spectra
and polarization based
on the model described in the previous section.
We inject $N=3\times 10^8$ photon packets
in each calculation.
Note that the
setup of the calculation 
for the uniform jet (\S\ref{uniform}) and two-component jet model (\S\ref{Two})
are basically identical to the ones adopted in Paper I.
As mentioned in the previous section, 
Main difference
is that here we calculate
the polarization state of the photon
and its effect on the scattering
 (this was neglected in the previous study).\footnote{
It is also noted that,
since the effect of the 
polarization on the overall track of the photon propagation is  
not significant, the obtained spectra are almost identical to
those obtained in the previous study.}
%

\subsection{Uniform (Non-Stratified) Jet}
\label{uniform}

Before considering a stratified jet, 
we first present results for a one-component uniform jet
that does not have structures in the $\theta$ direction
 ($\theta_0 = \theta_{\rm j} = 1^{\circ}$).
The isotropic equivalent kinetic luminosity, the dimensionless entropy
(terminal Lorentz factor) and the initial radius of the fireball
are set to be
$L_0=10^{53}{\rm erg~s^{-1}}$, 
$\eta_0=400$
and
$r_{\rm i}=10^8~{\rm cm}$, respectively.
Unpolarized photon packets ($Q=U=V=0$)
are injected at a radius 
$r_{\rm inj}= 4\times 10^{10}\eta_{0,400}r_{\rm i,8}~{\rm cm}$
with intensity given by a blackbody of
comoving temperature  
$k_{\rm B} T'_{\rm inj} = 
1.7r_{\rm i, 8}^{-1/2}\eta_{0, 400}^{-1}L_{0,53}^{1/4}~{\rm keV}$
 (see \S\ref{model} for detail), where
 $\eta_{0,400}=\eta_0/400$, $L_{0,53}=L_0/10^{53}~{\rm erg~s^{-1}}$ and
 $r_{\rm i,8}=r_{\rm i}/10^{8}~{\rm cm}$. 
%

\begin{figure*}[htbp]
\begin{center}
\includegraphics[width=17cm,keepaspectratio]{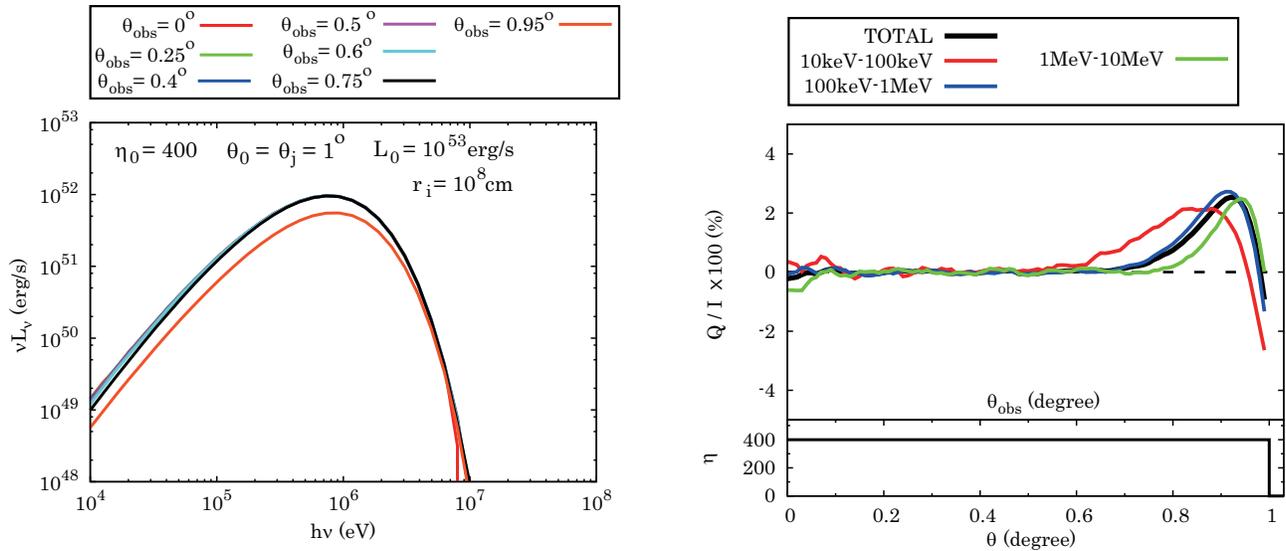}
\end{center}
\caption{
Left: Observed luminosity spectrum  for a uniform jet
 ($\theta_0 = \theta_{\rm j} = 1^{\circ}$)
  with
 parameters of the fireball
 given by $L_0=10^{53}~{\rm erg~s^{-1}}$, $\eta_0=400$ and
 $r_{\rm i}=10^8~{\rm cm}$.
 The various solid lines
 correspond to the different observer angles as shown in the legend.
Right: Polarization degree
({\it top}) and dimensionless entropy ({\it bottom})
 as a functions of  observer angle
and lateral position of jet, respectively.
The thick black solid line shows the result obtained by 
summing up all photons,
while the solid  red, blue and green lines correspond
to the photons in the energy bins of 
$10~{\rm keV}-100~{\rm keV}$,
$100~{\rm keV}-1~{\rm MeV}$ and
$1~{\rm MeV}-10~{\rm MeV}$, respectively.
The dashed line indicates $Q/I=0$ for reference.
Here and in subsequent plots,
we do not include the viewing angle bins that have
 less than $2000$ photon packets to avoid low statistics.
We also neglect the observers that have LOS 
located outside the lateral jet boundary
($\theta_{\rm obs}>\theta_{\rm j}$) as mentioned in \S\ref{Eval}.
}
\label{UNI}
\end{figure*}

In Fig.~\ref{UNI}, we display the obtained numerical results.
%
%
The left panel shows the spectra
for various observer angles $\theta_{\rm obs}$.
%
%
The obtained spectra have thermal shape due to the absence of sharp structures
inside the jet.
For an observer located at $\theta_{\rm obs}\lesssim \theta_{\rm j}- \eta_0^{-1} \sim 0.86^{\circ}$,
the results do not vary from those obtained in the case for spherical 
outflow, since the effect of the lateral jet boundary is not significant 
for the  
photons located at $\theta \sim \theta_{\rm j}- \eta_0^{-1} \sim 0.86^{\circ}$
due to the relativistic beaming effect.
Hence, observed spectra are
almost identical at this range of the observer angle.
The peak energy and luminosity can be estimated as
$E_{\rm p}\sim
 800 r_{\rm i,8}^{1/6}\eta_{400}^{8/3}L_{53}^{-5/12}~{\rm keV}$
and 
$L_{\rm p} \sim  10^{52} r_{\rm i,8}^{2/3}\eta_{400}^{8/3}L_{53}^{1/3}~{\rm erg~s^{-1}}$, respectively
(see Paper I for detail).
At a larger observer angle
($\theta_{\rm obs}\gtrsim \theta_{\rm j}- \eta_0^{-1} \sim 0.86^{\circ}$),
while the shapes of the spectra remain nearly unchanged, instead
the luminosity  decreases as the observer angle increases.
This is simply because 
a fraction of the cone of 
half opening angle $\sim \eta_0^{-1}$ around LOS 
falls out the lateral jet boundary ($\theta > \theta_{\rm j}$).
%
%
%
The obtained  spectra are  broader than 
the Planck function.
Somewhat softening is seen in below the peak energy,
resulting in the spectra $\nu L_{\nu} \propto \nu^{2}$,
in contrast with the Planck one $\nu L_{\nu} \propto \nu^{3}$.\footnote{In Paper I,
although the obtained spectra were identical,
the evaluation of low-energy spectral slope was 
not  accurate  and indicated in the text as
$\nu L_{\nu} \propto \nu^{2.4}$.
It is noted that the estimation in the present paper ($\nu L_{\nu} \propto \nu^{2}$)
is more accurate and is in agreement with 
other studies \citep[e.g.,][]{BSV13}.}
This is due the contribution of regions off-aligned from the LOS
that have lower thermal peak energies 
due to the smaller Doppler factor.

The right panel of  Fig.~\ref{UNI} shows the dependence of 
polarization signal $Q/I$ as a function of the observer angle.
The thick black solid line shows the result obtained by 
summing up the total photons,
while the 
red, 
blue
and green
 solid lines correspond
to the cases for the photons in the  limited frequency bins of 
$10~{\rm keV}-100~{\rm keV}$,
$100~{\rm keV}-1~{\rm MeV}$ and 
$1~{\rm MeV}-10~{\rm MeV}$, respectively.
%
%
As mentioned in \S\ref{Eval}, 
break of axisymmetry in the emission region around
the LOS  within an angle $\sim \eta_0^{-1}$
is required to produce non-zero polarization degree.
In the case of the uniform jet, the origin of the 
asymmetry in the emission region is 
solely due to the presence of the jet edge.
Therefore,
the polarization degree 
equals zero when the observer angle is well below
$\theta_{\rm j}-\eta_0^{-1}\sim 0.86^{\circ}$,
since the effect of the jet edge is negligible.
Although small ($|Q|/I \lesssim 3\%$),
non-zero polarization degree is found at a larger observer angle.
%
Regarding the dependence of $Q/I$ on the observer angle,
it increases  with the observer angle, and 
then drops to zero or negative at  $\theta_{\rm obs}\sim \theta_{\rm j}$.

%

The behaviour of the polarization
at observer angles near to the jet edge
 can be explained as follows:
as mentioned in \S\ref{Eval}, 
when the observer angle is close to but below the jet
opening angle $\theta_{\rm obs} \lesssim \theta_{\rm j}$,
a fraction of the cone of 
half opening angle $\sim \eta_0^{-1}$ around LOS
falls out the lateral jet boundary ($\theta > \theta_{\rm j}$).
First, as the observer angle increases,
the emission regions that produce negative $Q$ photon fluxes
(electric vector of polarization 
 perpendicular to the plane formed by
 the LOS and the jet axis) 
are cut away.
Therefore,
the observed net polarization have
a positive $Q$.
As the observer angle
becomes  much closer to the 
jet edge, also the regions emitting photons having  
 positive $Q$ begins to be cut away.
Hence, it leads to the decrease in $Q/I$.
The photon fluxes with
 positive and negative $Q$ nearly balance 
 at $\theta_{\rm obs} \sim \theta_{\rm j}$
to produce unpolarized signature.
%

Regarding the dependence on the frequency,
the increase and decrease in $Q/I$ tend to 
appear at smaller $\theta_{\rm obs}$
for low energy photons  ($10~{\rm keV}-100~{\rm keV}$) 
than those for higher energy photons.
This is due to the fact that the
photon with energies far below the peak
is produced mainly at the regions
more off-aligned from the LOS than those with higher energies.
The contribution of off-aligned component becomes important at 
low energies because the Doppler factor is smaller.

\subsection{Two-component Jet}
\label{Two}
\subsubsection{Infinitesimal boundary width $d\theta_{\rm B}=0$}
\label{Twoin}

\begin{figure*}[htbp]
\begin{center}
\includegraphics[width=17cm,keepaspectratio]{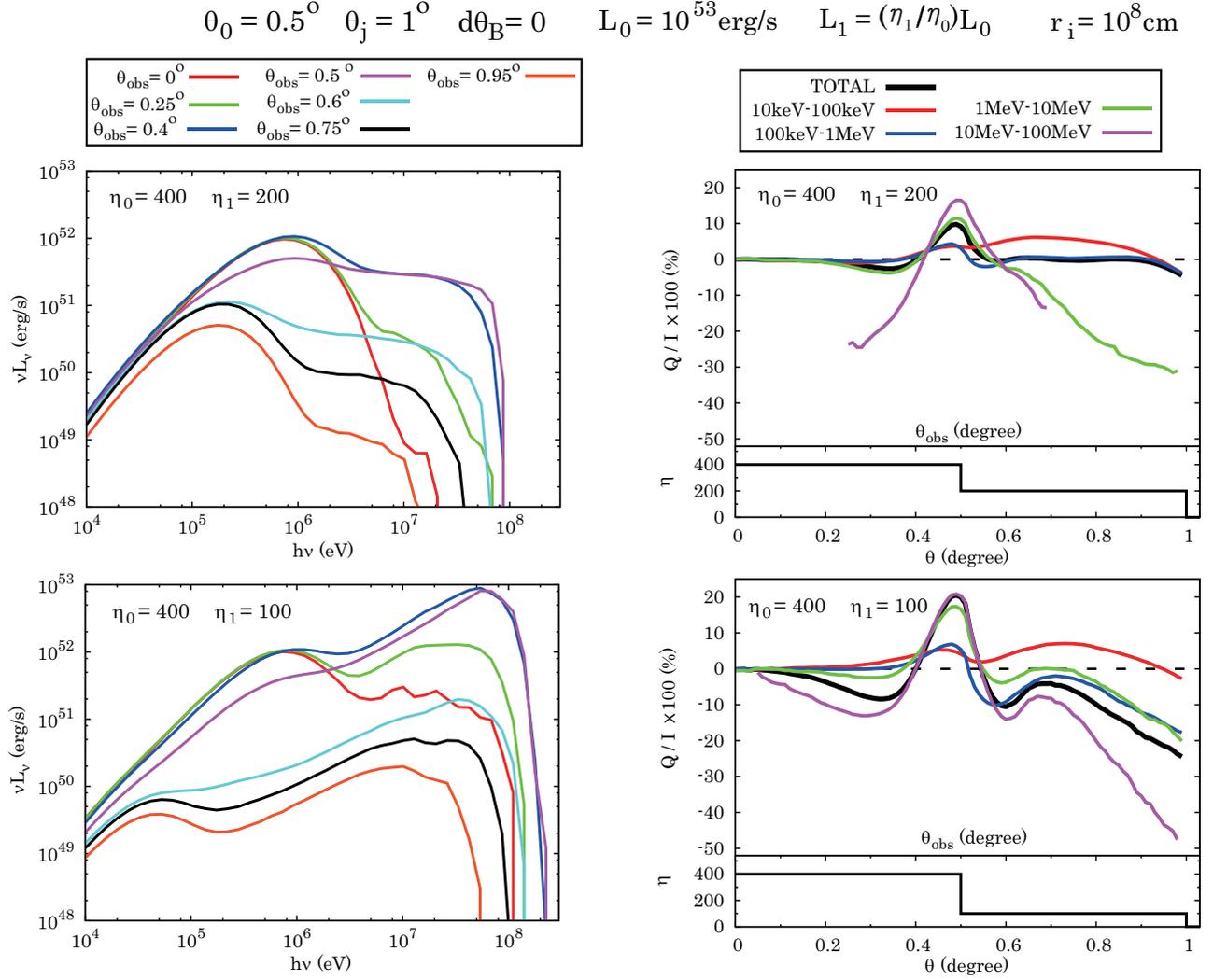}
\end{center}
\caption{
Same as Fig.~\ref{UNI}, but 
for the case
 of a two-component (spine-sheath) jet
 in which the spine jet with half opening angle of $\theta_0=0.5^{\circ}$
 is embedded in a wider sheath outflow with half opening angle of
 $\theta_{\rm j}=1^{\circ}$.
 The width of the transition boundary layer is set to be 
 infinitesimal ($d\theta_{\rm B}=0$).
The top panels show
the case for sheath region with 
 $\eta_1=200$ and $L_1=(\eta_1/ \eta_0)L_0=5\times10^{52}~{\rm erg~s^{-1}}$,
 while bottom panels show the case for 
 $\eta_1=100$ and $L_1=(\eta_1/ \eta_0)L_0=2.5\times10^{52}~{\rm erg~s^{-1}}$.
 Fixed values are employed
 for dimensionless entropy (terminal Lorentz factor)
 and kinetic luminosity in the spine region in both cases 
 ($\eta_0=400$ and  $L_0=10^{53}~{\rm erg~s^{-1}}$).
 The initial radius of the fireball is  
 $r_{\rm i}=10^8~{\rm cm}$ in both regions. 
}
\label{400-200_100}
\end{figure*}

 Here we show the results for a two-component stratified jet
 model in which the width of the 
 boundary transition layer is infinitesimal ($d\theta_{\rm B}=0$).
 We consider two cases
 where the
 imposed fireball parameters
 in the spine region  are identical,
 but
 those in the sheath regions are different.
As for the spine region,
we employ the same parameter set 
assumed in the uniform jet model
($\eta_0=400$ and $L_0=10^{53}~{\rm erg~s^{-1}}$).
As for the sheath regions, 
we consider the cases with
 $\eta_1=200$ ($L_1=(\eta_1/ \eta_0)L_0=5\times10^{52}~{\rm erg~s^{-1}}$)
and
 $\eta_1=100$ ($L_1=(\eta_1/ \eta_0)L_0=2.5\times10^{52}~{\rm erg~s^{-1}}$).
%
%
%
In the top and bottom panels of Fig.~\ref{400-200_100},
 we display
 obtained numerical results
 for the former and the latter case, respectively.
%
 The left and right panels of the figure display
 the  observed spectra and the polarization, respectively. 
 In both cases, the
 initial fireball radius,  half opening angle of the spine
 and the whole jet are fixed at
 $r_{\rm i}=10^{8}~{\rm cm}$,
 $\theta_0 = 0.5^{\circ}$ and $\theta_{\rm j} = 1^{\circ}$, respectively.
%

 As seen in the figure,
 the appearance of the spectrum
 deviates significantly from a thermal one.
 Above the thermal peak,
 population of photons that gained energy
by  crossing the boundary layer multiple times
 produce a non-thermal tail in the spectrum (see Paper I). 
%
 The efficiency of the photon acceleration increases
 as the  difference in the velocity (Lorentz factor) 
 between the two regions becomes larger.
 Hence,
 the non-thermal component in the spectra
 is harder for larger velocity difference.
%
%
 The photon acceleration becomes inefficient when the
 photon energy becomes large enough so that the
 recoil
 of electrons cannot be neglected (Klein-Nishina effect).
 Hence, in all cases, the spectrum does not extend up to energies higher than
 $h\nu \sim \Gamma_0 m_e c^2 \sim 200(\Gamma_0/400)~{\rm MeV}$.
 This implies that
 our model predicts a high energy cut-off
 around $\sim 100~{\rm MeV}$, when 
 the bulk Lorentz factor of the jet is $\sim {\rm a~ few}~100$.
%
%

%
 The observed spectrum is sensitive to the  observer angle. 
 The non-thermal tail is
 hardest 
 when the LOS is aligned to the 
 boundary layer $\theta_{\rm obs}=\theta_0 = 0.5^{\circ}$
 and becomes softer as the deviation between 
 $\theta_{\rm obs}$ and $\theta_0$ becomes larger, simply
 because the boundary layer corresponds to the site of 
 photon acceleration.

 The thermal peak energy and 
 the luminosity also change
 with the observer angle.
%
 For an observer at $\theta_{\rm obs} < \theta_0$, the 
 thermal component is determined mainly by 
 photons which have propagated through the spine region.
 Therefore,
 the peak energy and luminosity are
 roughly equal to the case of the uniform jet
 considered above 
 in which a same set of parameters
 ($\eta_0$, $L_0$ and $r_{\rm i}$) is assumed.

 On the other hand,
 for an observer at $\theta_{\rm obs} > \theta_0$,
 photons which have propagated through the sheath region 
 dominate the thermal component.
 Accordingly, the thermal peak energy 
 and luminosity are roughly
 lower by a factor $\sim (\eta_0/\eta_1)^{8/3}(L_0/L_1)^{-5/12} = (\eta_0/\eta_1)^{27/12}\sim 4.7(\eta_{0,400}/\eta_{1,200})^{27/12}$
 and $\sim (\eta_0/\eta_1)^{8/3}(L_0/L_1)^{1/3}=(\eta_0/\eta_1)^3 
 \sim 8(\eta_{0,400}/\eta_{1,200})^3$, respectively, where
 $\eta_{1,200}=\eta_1/200$.
 It is noted, however, that the peak luminosity 
 can be lower by several factors
 when the observer angle is near the jet  edge ($\theta_{\rm obs} > \theta_{\rm j} - \eta_1^{-1}$).
This is due to the fact that
 a cone of half-opening angle
 $\sim \eta_1^{-1} =0.57^{\circ}(\eta_{1,100})^{-1}$
  around the LOS,
 in which most of the observable emissions are produced,
 exceeds the jet region,
 where
 $\eta_{1,100}=\eta_1/100$.
%

 It is worth noting that
 the  spectra below the thermal peaks
 are slightly softened, compared to those observed in the case
 of uniform jet $\nu L_{\nu} \propto \nu^{2}$.
 This is mainly due to
 the multi-color temperature effect.\footnote{Although we did not mention explicitly in the text,
 this effect
 has already been observed
 in the result presented in Paper I.}
%
 As mentioned in \S\ref{uniform}, at low energies,
 contribution of regions off-aligned from the LOS becomes significant.
 When the LOS of the observer lying within the spine (sheath) region
 the sheath (spine) region
 contaminates to the off-aligned region.
 Since each region has a
 different thermal peak energy, luminosity and Doppler factor,
 superposition of these emissions leads to
 a slight softening, 
 resulting in  $\nu L_{\nu} \propto \nu^2$-$\nu^{1.5}$ is observed.
%


 As seen in the right panels of Fig.~\ref{400-200_100},
 the polarization signal
 is more pronounced than the case of the uniform jet.
%
 In the case of the two-component jet,
 in addition to the edge of the jet, the sharp boundary 
 layer between the spine and sheath produces another site for
 the break of axisymmetry around the LOS.
%
 The significant change in the emission properties
 around the LOS
 produces a prominent peak in the distribution of 
 polarization signal for the overall emission
 ($h\nu = 10~{\rm keV}-100~{\rm MeV}$: {\it Black lines}),
 at $\theta_{\rm obs}\sim \theta_0$.
%
 The  polarization degree is larger when the difference in the  
 velocity (Lorentz factor) is larger,
 since the change in the emission is more prominent.
In the case of $\eta_0=400$ and $\eta_1=100$,
 the peak polarization degree can be as high as $\sim 20\%$.

\begin{figure}[ht]
\begin{center} 
\includegraphics[width=8.5cm]{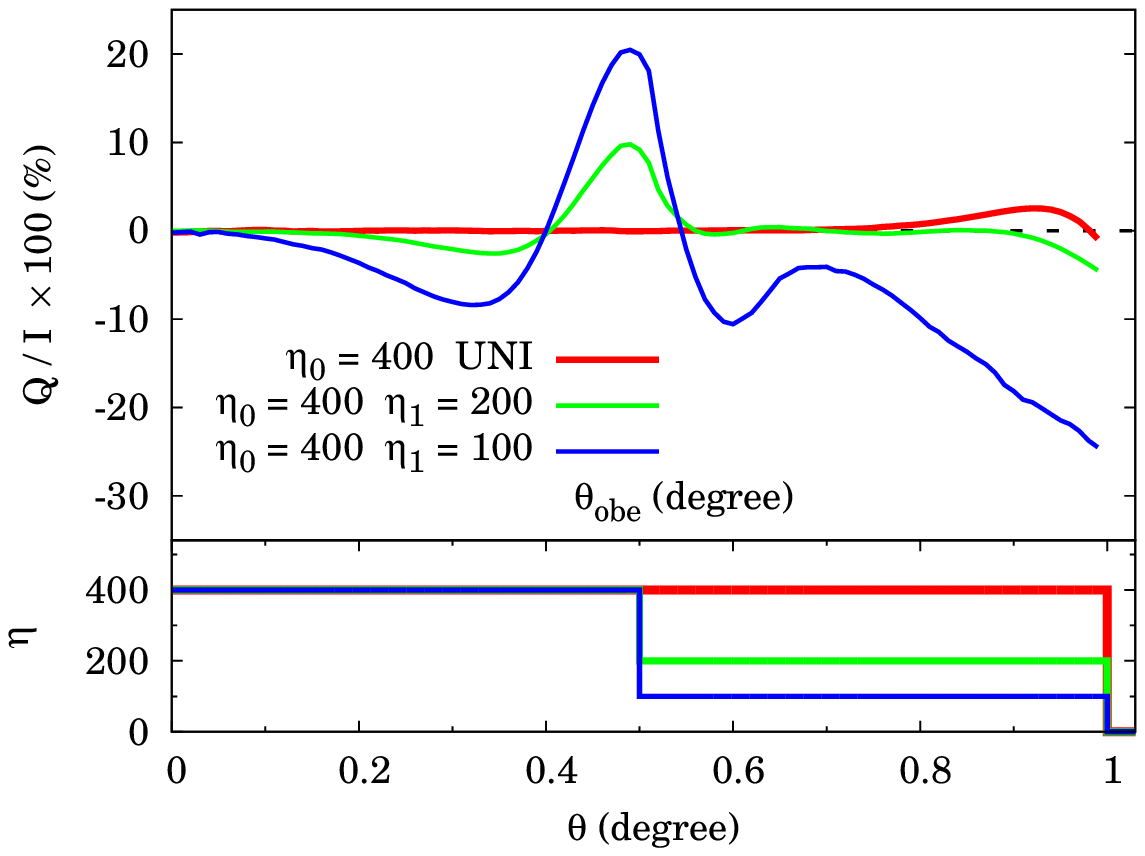}
\end{center}
\caption{
 Transverse distribution of the dimensionless entropy ({\it bottom}) 
 and  
total polarization degree as a function
 of observer angle ({\it top})  
 for a uniform jet  ($\theta_0=\theta_{\rm j}=1^{\circ}$)  and a
 two-component jet
 in which the spine jet with half opening angle of $\theta_0=0.5^{\circ}$
 is embedded in a wider sheath outflow with half opening angle of
 $\theta_{\rm j}=1^{\circ}$.
 The employed
 values for the dimensionless entropy (terminal Lorentz factor)
 and kinetic luminosity 
 for the uniform jet and the spine region of the two-component jet 
 are fixed
 as $\eta_0=400$ and  $L_0=10^{53}~{\rm erg~s^{-1}}$, respectively.
 The initial radius of fireball is fixed as  
 $r_{\rm i}=10^8~{\rm cm}$ in all cases.
 The red solid line shows the case of a uniform jet, while
 the green and blue solid lines show
 the cases
 for the two-component jet
 having dimensionless entropies
 given by  $\eta_1=200$ and  $\eta_1=100$
 in the sheath region, respectively.
 In each case, the kinetic luminosity in the sheath region is
 given by $L_1=(\eta_0/\eta_1)L_0$.
%
}
\label{Twocomp}
\end{figure}

 To clarify how the presence
 of the velocity shear and its amplitude
 affects the resulting polarization,
  we also show 
 comparison of the 
 assumed transverse distribution  of 
 the dimensionless entropy as well as the
 corresponding observed total polarization degree 
 among the uniform jet and two-component jet models in Fig.~\ref{Twocomp}.
%
 
%

 In the case of a smaller Lorentz factor difference
 ($\eta_0=400$ and $\eta_1=200$), 
 the emission is brightest near the thermal peak energies.
%
 On the other hand, in the case of larger Lorentz factor difference
 ($\eta_0=400$ and $\eta_1=100$), 
 non-thermal emissions above the thermal peak energies
 are the brightest in most of the observer angles. 
 Hence, the overall polarization
 traces the behaviour of the polarization near
 the thermal peak
 ($h\nu\sim 1~{\rm MeV}$ for  $\theta_{\rm obs} \lesssim \theta_0=0.5^{\circ}$ and
 $h\nu\sim 200~{\rm keV}$ for $\theta_{\rm obs} \gtrsim \theta_0=0.5^{\circ}$)
 for the former case, 
 while, in the latter case, polarization signature of the non-thermal
 emissions
 ($h\nu> 1~{\rm MeV}$ for  $\theta_{\rm obs} \lesssim \theta_0=0.5^{\circ}$ and
 $h\nu > 100~{\rm keV}$ for $\theta_{\rm obs} \gtrsim \theta_0=0.5^{\circ}$)
 are traced.

 Since the photon acceleration site 
 is localized at a certain lateral position of the jet $\theta \sim \theta_0$,
 the properties of the non-thermal emission change 
 more sensitively with lateral angle $\theta$ than
 the thermal emissions.
 Consequently, the polarization signal
 of the non-thermal emissions
 tends to be stronger than that of the  thermal emissions
 and, therefore,
 overall polarization signal is stronger in the latter case
 than the former case.
 The polarization degree of the non-thermal emissions 
 can be as high as $\sim 30 - 40 \%$,
 when the observer is located near the edge of the jet 
 $\theta_{\rm obs} \sim \theta_{\rm j}$
 (for example, see 
 green and magenta line
 in the top and bottom panels of Fig.~\ref{400-200_100}, respectively).

 The  observer dependence of
 the overall polarization signal $Q/I$ can be
 explained as  follows :
 regarding the lateral $\theta$ dependence of the emitting region,
 the luminosity of the overall emission increases with 
 lateral position when $\theta \leq\theta_0=0.5^{\circ}$,
 due to the appearance of the non-thermal photons.
 %
%
 Hence, when the observer angle is 
 in the range
 $\theta_{\rm obs} \lesssim \theta_0-\eta_0^{-1} \sim 0.36^{\circ}$,
 within the beaming cone 
 around the LOS,
 the luminosity increases with $\theta$.
 This implies that photons flux possessing negative $Q$
 is strongest within the cone.
 As a result,
 net polarization signal $Q/I$ becomes
 negative.
 At a larger observer angle
  $\theta_0-\eta_0^{-1} \sim 0.36^{\circ} \lesssim
   \theta_{\rm obs} \lesssim \theta_{0}=0.5^{\circ}$,
 sheath region enters within the beaming cone.
 Since the sheath region is much dimmer than the spine region,
 this leads to the decrease in 
 the photon flux with negative $Q$. 
 Therefore, 
 polarization signal $Q/I$  increases in the positive direction
 as $\theta_{\rm obs}$ increases, and reaches maximum at
 $\theta_{\rm obs}\sim \theta_0$.
 Then, at $\theta_{\rm obs} \gtrsim \theta_0$,
 the polarization  $Q/I$ decreases rapidly with $\theta_{\rm obs}$
  and again becomes negative,
 since the emission region producing  photon flux with positive $Q$ decreases.
 This rapid decrease ceases at an observer angle below
 $\sim \theta_0 + \eta_0^{-1}\sim 0.64^{\circ}$
 when the photon flux from the sheath region becomes comparable to 
 that from the spine region.
 At  $\theta_0 + \eta_0^{-1}\sim 0.64^{\circ} \lesssim \theta_{\rm obs}$,
 photons from the spine region become negligible, and  
 the observable photon flux is dominated by those emitted from the sheath 
 region.
 Hence, when the total flux is dominated by the thermal photons 
 as in the case of $\eta_0=400$ and $\eta_1=200$,
 overall polarization signal becomes weak
 since 
 the luminosity of the emitting region is not so sensitive  to the lateral position.
 On the other hand, when the 
 total flux is dominated by the non-thermal photons
  as in the case of $\eta_0=400$ and $\eta_1=100$,
 the emitted photons are concentrated at the region near the
 boundary $\theta \sim \theta_0$, and decreases 
 rapidly with $\theta$.
 As a result, the photon flux that posses negative $Q$
 becomes pronounced as $\theta_{\rm obs}$ increases, which 
 in turn leads to decrease of $Q/I$.

\subsubsection{Non-zero boundary width $d\theta_{\rm B}>0$}
\label{Twofin}

 Here we show the results for a two-component stratified jet
 model in which the 
 boundary transition layer has a  finite width ($d\theta_{\rm B}>0$).
 The set up of our calculation is 
 basically the same as the case of higher velocity difference 
 considered in the previous section (\S\ref{Twoin}).
 In all cases,
 we employ the same fireball parameters 
 ($\eta_0=400$, $\eta_0=100$, $L_0=10^{53}~{\rm erg~s^{-1}}$,  
  $L_1=2.5 \times 10^{52}~{\rm erg~s^{-1}}$ and $r_{\rm i}=10^8~{\rm cm}$)
 and 
 the midpoint of the spine-sheath boundary layer
 and the half-opening angle of the jet are fixed as 
 $\theta_0 = 0.5^{\circ}$ and $\theta_{\rm j} = 1^{\circ}$, respectively.
 The only difference is that the boundary width 
 is finite and all the physical properties within the
 boundary layer is determined by linear interpolation 
 as explained in \S\ref{TrS}.

\begin{figure*}[htbp]
\begin{center}
\includegraphics[width=15.5cm,keepaspectratio]{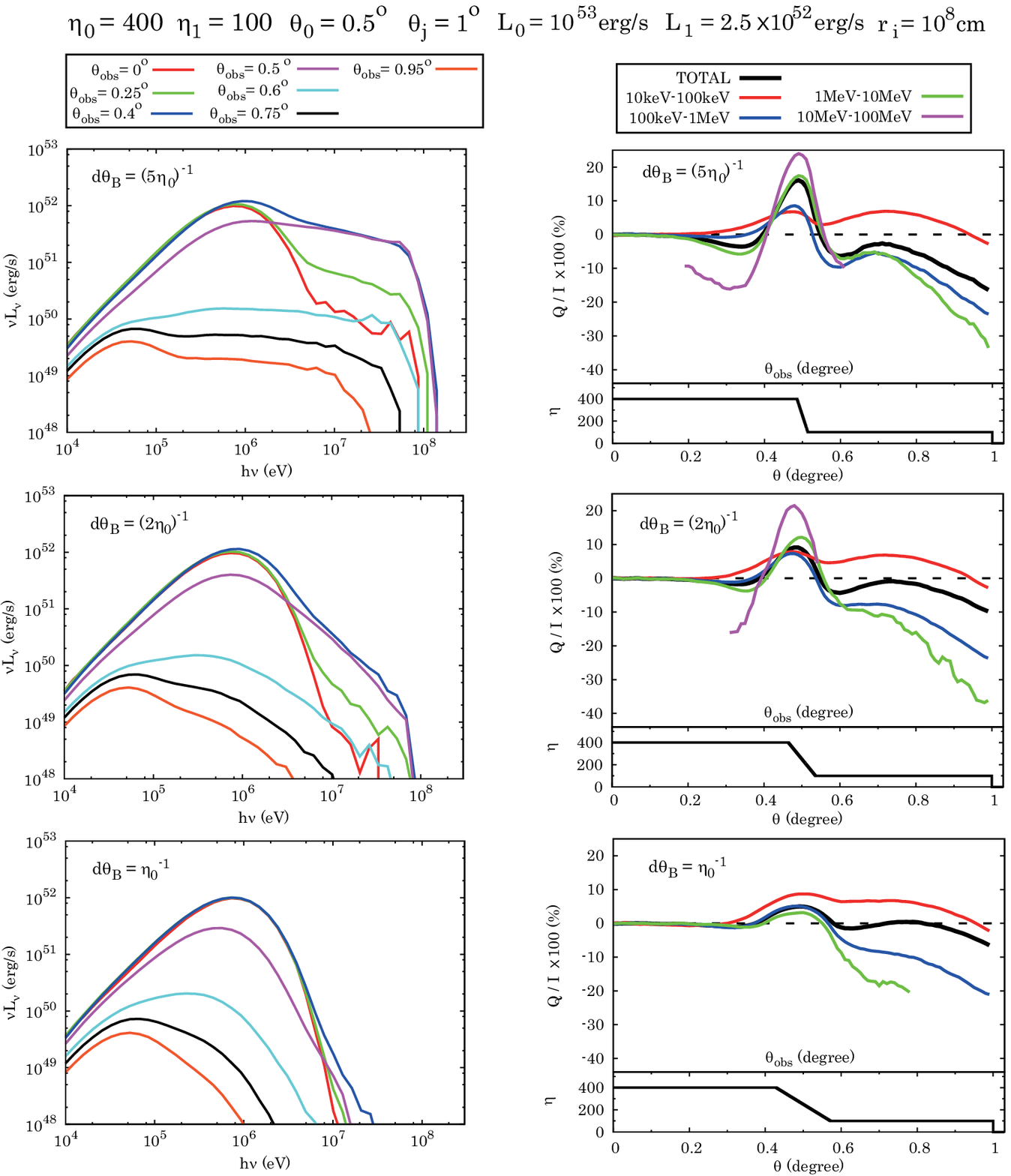}
\end{center}
\caption{
Same as bottom panels of Fig.~\ref{400-200_100}, but for the cases of  a
finite boundary width:
$d\theta_{\rm B}= (5\eta_0)^{-1}$ ({\it top}),
$(2\eta_0)^{-1}$ ({\it middle}) and
$\eta_0^{-1}$ ({\it bottom}).
}
\label{sm400-100}
\end{figure*}

 As for the widths of the boundary layer,
 three cases, 
 $d\theta_{\rm B}=(5\eta_0)^{-1}\sim 0.029^{\circ}$,
 $d\theta_{\rm B}=(2\eta_0)^{-1}\sim 0.072^{\circ}$ and
 $d\theta_{\rm B}=\eta_0^{-1}\sim 0.14^{\circ}$,
 are considered. 
 The numerical results are displayed in 
 Fig.~\ref{sm400-100}. 
 While the left panels of the figure show the 
 spectra for a given observer position,
 the right panels show the observer dependence of
 the polarization signal $Q/I$.
 Compared with the case of infinitesimal width
 (bottom panel of Fig.~\ref{400-200_100}),
 while the spectra up to the thermal peak energy do not show significant difference,
 spectra at higher energies which are dominated by the non-thermal photons 
 become much softer in this case.
 The non-thermal spectra becomes softer as the boundary layer becomes wider.
 This is simply because the efficiency of the  photon acceleration
 becomes lower when the gradient of the Lorentz factor is smaller
 (for detail, see Paper I).

 The general feature of the polarization signal does not vary 
 from the case of spine-sheath jet having a boundary
 layer of infinitesimal width.
 Regarding the observer angle dependence, 
 there is a characteristic peak at
 $\theta_{\rm obs}\sim \theta_0=0.5^{\circ}$. 
 As in the  case shown in the previous section,
 polarization signal is mainly governed by  
 the non-thermal photons that are produced at the boundary layer.
 The contribution of the non-thermal photons 
 in the total emission becomes smaller
 as the boundary layer becomes wider.
 since the non-thermal photons becomes less pronounced.
 Therefore,
 the overall polarization degree
 ($h\nu = 10~{\rm keV}-100~{\rm MeV}$)
 is lower in the cases with wider boundary widths.
%
 It should be noted, however, 
 that 
 the non-thermal photons themselves 
 show quite high polarization degree ($\gtrsim 20\%$)
 even in the cases for weak non-thermal emissions
 (e.g., see green lines of Fig.~\ref{sm400-100}).

\begin{figure}[ht]
\begin{center} 
\includegraphics[width=8.5cm]{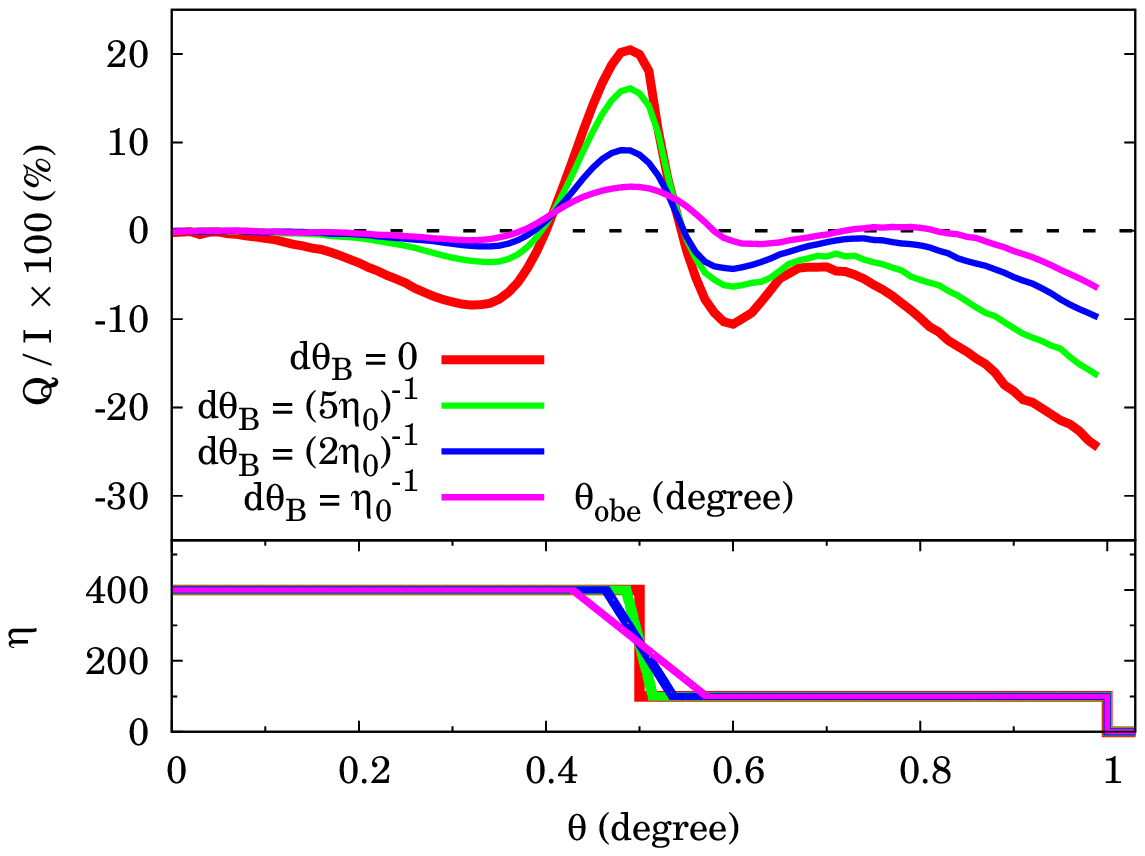}
\end{center}
\caption{
 The transverse distribution of the dimensionless entropy
 ({\it bottom panel}) and
total polarization degree as a function
 of observer angle ({\it top panel}) for
a two-component jet model in which
the  spine and sheath regions are 
located in the transverse range
$0\leq \theta \leq \theta_0 - d\theta_{\rm B}/2$ and 
$\theta_0 + d\theta_{\rm B}/2 \leq \theta \leq \theta_{\rm j}$,
respectively.
The figure shows the
case for  $\theta_0 = 0.5^{\circ}$ and  $\theta_{\rm j} = 1^{\circ}$.
The employed
values for dimensionless entropy (terminal Lorentz factor)
and kinetic luminosity are chosen
as $\eta_0=400$ and  $L_0=10^{53}~{\rm erg~s^{-1}}$ for the spine and
 $\eta_1=100$ and  $L_1=2.5\times 10^{52}~{\rm erg~s^{-1}}$ for the sheath.
The corresponding quantities in the boundary transition regions
are determined by the linear interpolations from the two regions.
 The initial radius of fireball is chosen as  
 $r_{\rm i}=10^8~{\rm cm}$ in all regions.                                                                              
 The solid red, 
green,
blue and 
magenta  lines
show the cases for the boundary layer widths of
$d\theta_{\rm B}=0$,
$d\theta_{\rm B}=(5\eta_0)^{-1}$,
$d\theta_{\rm B}=(2\eta_0)^{-1}$ and
$d\theta_{\rm B}=\eta_0^{-1}$, respectively.
}
\label{Twocompsm}
\end{figure}

 In Fig.~\ref{Twocompsm},
 to clarify the relation between the 
 boundary width and the polarization signal,
 we show the dimensionless entropy (terminal Lorentz factor)
 as a function lateral position $\theta$ (bottom panel)
 and the overall polarization signal $Q/I$ as a function 
 of $\theta_{\rm obs}$ (top panel)
 for the cases of
 $d\theta_{\rm B}=0$,
 $d\theta_{\rm B}=(5\eta_0)^{-1}$,
 $d\theta_{\rm B}=(2\eta_0)^{-1}$ and
 $d\theta_{\rm B}=\eta_0^{-1}$.
 Indeed, it is seen that the overall polarization signal 
 is weaker for the cases with wider boundary widths.

\subsection{Multi-component Jet}
\label{Multi}

 Here we show the results for a stratified jet in which multiple-components
 are present. 
 Let us clarify again our motivation 
 for considering such a structure.

 In the previous sections, we have shown that
 the emissions from
 two-component
 jet that has strong velocity gradient 
 show non-thermal spectra  and accompany high polarization degree.
 As  seen in Figs.~\ref{400-200_100} and \ref{sm400-100}, 
 by adopting appropriate values for our parameter set,
 the two-component jet can reproduce the 
 observed spectra of GRBs.
%
%
 For example, focusing on the case shown in the left top
 panel of  Fig.~\ref{sm400-100} 
 ($\eta_0=400$, $\eta_1=100$ and $d\theta_{\rm B}=(5\eta_0)^{-1}$),
 the spectra below and above the thermal peak energy
 that are located at $\sim 800~{\rm keV}$ 
 can be roughly approximated as 
 $\nu L_{\nu} \propto \nu^{1.5}$ and
 $\nu L_{\nu} \propto \nu^{-0.5}$-$\nu^{-0.3}$, respectively,
 for an observer nearly aligned to the boundary layer
 $\theta_{\rm obs}\sim 0.4-0.5^{\circ}$.

 These features are consistent with the observations
 that are often modeled by a Band function which shows
 a smoothly joined broken power-law that peaks at $\sim$ a few $100~{\rm keV}$.
 The photon indices below ($\alpha_{\rm ph}$) and above ($\beta_{\rm ph}$)
 the peak energy
 vary
 from source to source
 ($\nu L_{\nu}\propto \nu^{\alpha_{\rm ph} + 2}$ for $\nu < \nu_{\rm p}$
  and $\nu L_{\nu}\propto \nu^{\beta_{\rm ph} + 2}$ for $\nu > \nu_{\rm p}$),
 but have  typical values
 at $\alpha_{\rm ph}\sim -1$ and $\beta_{\rm ph}\sim -2.5$, respectively.
 Hence,
 the high energy slope matches the 
 typically observed  value.
 Although  the low energy slope is
 relatively hard, it
 is fairly close to the typical value ($\nu L_{\nu} \nu$)
 and is in the range of the observation \citep[e.g.,][]{NGG11, GBP12, GPM13}.

 However,
 for an observer far off-aligned from the boundary layer,
 the spectra depart from the typical values.
 This is due to the fact that the
 acceleration site of the photons is located at
 a single fixed lateral position $\theta =\theta_0$,
 which suppresses the accelerated photons 
 from spread out in 
 various direction due to the relativistic beaming.
%

 One possible solution to overcome this difficulty is to consider
 a presence of velocity shear in various lateral position 
 of the jet.
%
If velocity shear regions are distributed 
within the entire jet 
 (closely spaced within an angular scale $\sim 2\Gamma^{-1}$),
the acceleration photons will be prominent 
for all observers.
%
%

 Indeed, large number of 
 hydrodynamical simulations of axisymmetric
 jet propagation 
 show that large velocity gradient regions, such as those
 accompanied by recollimation shocks, actually 
 appear across the whole jet
 \citep{ZWM03, MYN06, MLB07, LMB09, MNA11, NIK11,MI13}. 
 Moreover, beyond the axisymmetry, we emphasize that
 hydrodynamical instabilities such as Rayleigh-Taylor
 and Richtmyer-Meshkov instabilities
 lead to appearance of multi-component jet having various 
 Lorentz factor.  
 This is  shown in the recent numerical simulations by \citet{MM13a, MM13b}.
 Their results indicated that, as the jet propagate
 through a dense medium,
 these instabilities produce small scale filamentary structures
 that have sharp interface to distribute in entire jet regions.
 This rich internal structure within the jet can 
 provide acceleration site in various angular scales and remove the 
 strong observer dependence.

\begin{figure*}[htbp]
\begin{center}
\includegraphics[width=15.5cm,keepaspectratio]{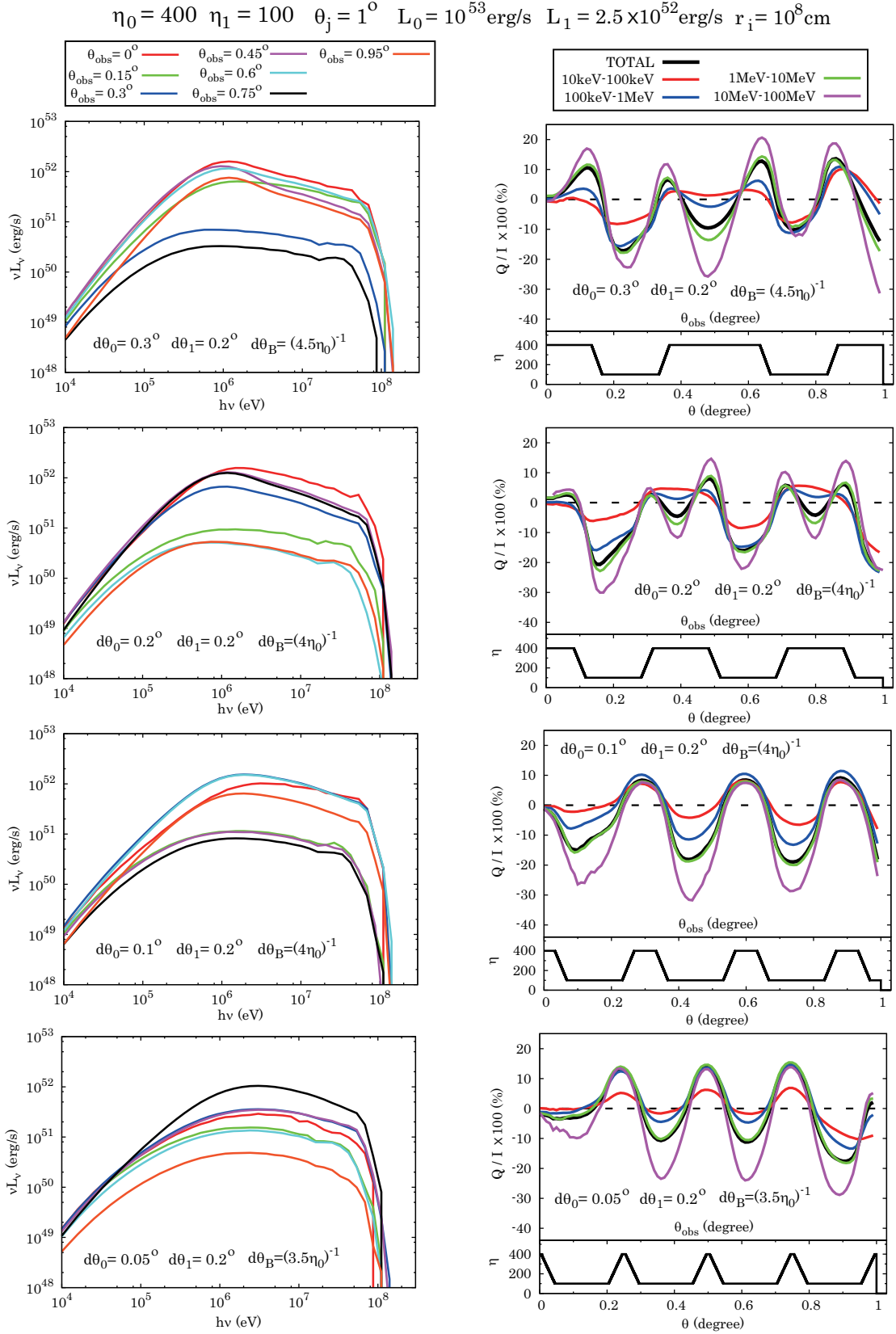}
\end{center}
\caption{
Same as Fig.~\ref{400-200_100}, but for the cases of  
multi-component jets.
Two components  
having fixed  widths of 
$d\theta_0-d\theta_{\rm B}$ (C0) and 
$d\theta_1-d\theta_{\rm B}$ (C1) alternately appear
in the transverse direction   
within the jet with half opening angle $\theta_{\rm j}=1^{\circ}$.
While $d\theta_1=0.2^{\circ}$ is employed in all cases,
from top to bottom panels, 
cases for $d\theta_0=0.3^{\circ}$,
$0.2^{\circ}$,
$0.1^{\circ}$ and 
$0.05^{\circ}$
are shown.
%
The employed
values for the dimensionless entropy (terminal Lorentz factor)
and kinetic luminosity are chosen
as $\eta_0=400$ and  $L_0=10^{53}~{\rm erg~s^{-1}}$ for C0 and
 $\eta_1=100$ and  $L_1=2.5\times 10^{52}~{\rm erg~s^{-1}}$ for C1 regions,
 respectively.
 The initial radius of fireball is chosen as  
 $r_{\rm i}=10^8~{\rm cm}$ in all regions. 
}
\label{m400-100}
\end{figure*}

\begin{figure*}[htbp]
\begin{center}
\includegraphics[width=15.8cm,keepaspectratio]{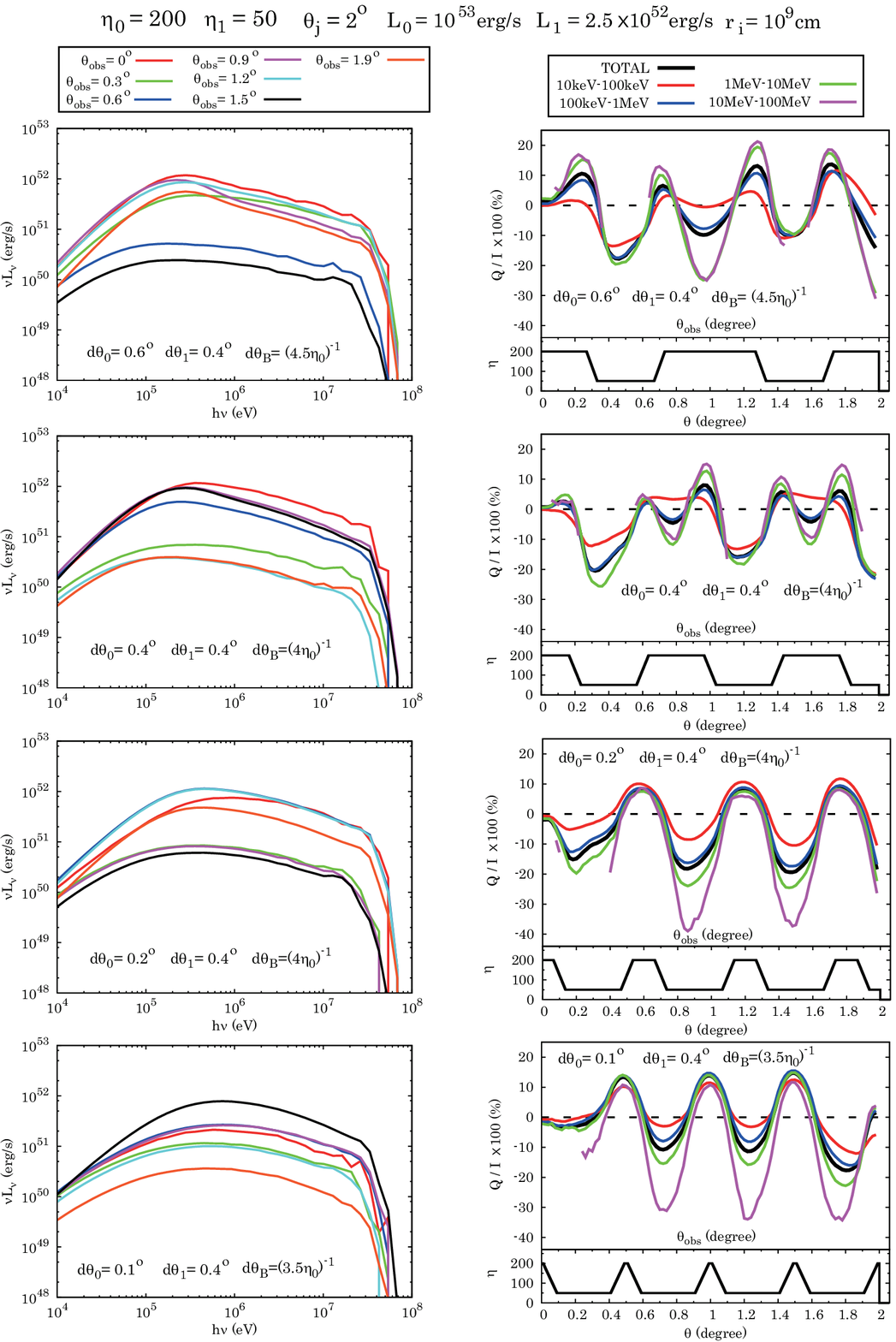}
\end{center}
\caption{
Same as Fig.~\ref{m400-100}, but for
lower dimensionless entropies
($\eta_0=200$ and $\eta_1=50$),  
wider 
jet
($\theta_{\rm j}=2^{\circ}$)
and larger initial fireball radius
($r_{\rm i}=10^9~{\rm cm}$).
While $d\theta_1=0.4^{\circ}$ is employed in all cases,
from top to bottom panels, 
case for $d\theta_0=0.6^{\circ}$,
$0.4^{\circ}$, 
$0.2^{\circ}$ and
$0.1^{\circ}$ 
are shown.
}
\label{m200-50}
\end{figure*}

 Motivated by this background,  we explore emissions from
 the jet having multiple components that have sharp velocity 
 gradients between their interfaces.
 We mimic the complex structure
 inferred from the simulations
 with a simplified jet structure that has a velocity 
 shear present at multiple lateral positions $\theta$
 as described in \S\ref{model}.
 Here, we focus on the 
 parameter ranges
 which result in spectra close to the typically observed ones.
%
%
 The obtained
 numerical  results
 (spectra and polarization signal $Q/I$)
 are presented in Figs.~\ref{m400-100}-\ref{m200-50}.
%
%

 In the cases shown in Fig~\ref{m400-100},
 the values of the fireball parameters
 imposed in the C0 and C1 regions are identical 
 to those employed in the spine and sheath regions
 of the two-component jet model with 
 larger difference in the dimensionless entropy, respectively
 ($\eta_0=400$, $\eta_1=100$, $L_0=10^{53}~{\rm erg~s^{-1}}$,
 $L_1=2.5\times 10^{52}~{\rm erg~s^{-1}}$ and $r_{\rm i}=10^8~{\rm cm}$).
 We also impose the same value for the half-opening angle of the jet
 ($\theta_{\rm j}=1^{\circ}$).
%
While
fixed value is employed for the 
angular extension of  C1 region
 ($d\theta_1=0.2^{\circ}$), 
from top to bottom panels, 
different values are employed for C0 region
($d\theta_0=0.3^{\circ}$,
$0.2^{\circ}$,
$0.1^{\circ}$ and 
$0.05^{\circ}$)

 On the other hand,   Fig.~\ref{m200-50}
 shows the cases for a
 factor $2$ lower values  in the  dimensionless entropy 
 ($\eta_0=200$, $\eta_1=50$, $L_0=10^{53}~{\rm erg~s^{-1}}$,
 $L_1=2.5\times 10^{52}~{\rm erg~s^{-1}}$ and $r_{\rm i}=10^9~{\rm cm}$).
 The
 angular extensions of the jet and each components
 are chosen 
 to be a factor $2$ larger than 
 the former cases
 (i.e., $\theta_{\rm j}=2^{\circ}$, $d\theta_1=0.4^{\circ}$
 and $d\theta_0 = 0.6^{\circ}$, $0.4^{\circ}$,
 $0.2^{\circ}$ and $0.1^{\circ}$),
 since the typical spreading angle of the 
 photons is larger by the same factor. 
%
%
 In any case,
 finite values are imposed in the widths of the boundary layer
 ($d\theta_{\rm B}=(4.5\eta_0)^{-1}$, 
 $(4\eta_0)^{-1}$ or $(3.5\eta_0)^{-1}$).


 As  seen in the figures, in all cases,
 the spectra have broad non-thermal shapes 
 irrespective to the observer angle.
 Prominent high-energy tail
 is always present,
 since the boundary layers
 are located within an angle
 $\Gamma^{-1} \sim 
 0.14^{\circ}\Gamma_{400}^{-1}$ from the LOS
 of any observer, where $\Gamma_{400}=(\Gamma/400)$.
 The observed high energy photon indices are roughly
 in the range $-2.5 \lesssim\beta_{\rm ph}\lesssim -2$.
 The non-thermal component shows a cut-off below
 $h\nu \sim \eta_0 m_e c^2 \sim 200  \eta_{0,400}~{\rm MeV}$,
 due to the Klein-Nishina effect. 
 Hence, the spectra extends up to higher energy 
 in the cases for high dimensionless entropies
 $\eta_0=400$  (Fig.~\ref{m400-100})
 than those for low dimensionless entropies  (Fig.~\ref{m200-50}).

%
 Since the photon flux from the C0 regions largely 
 exceeds those from the C1 regions at any observer angle,
 the peak energy is always roughly equal to
 the value expected from the fireball parameters 
 of the C0 region.
 In the cases for the high  
 and low dimensionless entropies,
 these values can be estimated as
 $E_{\rm p}\sim
 800 r_{\rm i,8}^{1/6}\eta_{0,400}^{8/3}L_{53}^{-5/12}~{\rm keV}$
  (Fig.~\ref{m400-100})
 and 
 $E_{\rm p}\sim
 180 r_{\rm i,9}^{1/6}\eta_{0,200}^{8/3}L_{53}^{-5/12}~{\rm keV}$  
 (Figs.~\ref{m200-50}),
 respectively, where
 $\eta_{0,200}=\eta_0/200$ and $r_{\rm i,9}=r_{\rm i}/10^9~{\rm cm}$.
 It is noted, however, that
 the effect of the  photon acceleration tends to
 shift the peak energy to a
 slightly higher value
 when the LOS is located within the C0 region.
%
 This effect is prominent,
 particularly in the cases with smaller $d\theta_0$,
 since the fraction of photons within the C0 component
 that experiences the acceleration increases.
 On the other hand, relatively lower 
 peak energy is found
 when the LOS is within the C1 region.
 This is because the 
 C0 regions are off-aligned, and, therefore, the
 photons
 emitted there have  
 lower Doppler factors than the aligned cases.

 The observed thermal peak luminosity  
 depends on the fraction of volume
 occupied by C0 region
 within 
 a  half-opening angle
 $\sim \eta_0^{-1}\sim 0.14^{\circ}\eta_{0,400}^{-1}$
 around LOS.
 When it is fully occupied, 
 the peak luminosity is roughly equal to 
 that expected from the 
 fireball parameters of the C0 region,
 which are 
 $L_{\rm p} \sim  10^{52} r_{\rm i,8}^{2/3}
 \eta_{0,400}^{8/3}L_{53}^{1/3}~{\rm erg~s^{-1}}$
  (Fig.~\ref{m400-100}),
 and 
 $L_{\rm p} \sim  8 \times 10^{51} r_{\rm i,9}^{2/3}
 \eta_{0.200}^{8/3}L_{53}^{1/3}~{\rm erg~s^{-1}}$
 (Fig.~\ref{m200-50})
 for the cases of  high 
 and low dimensionless entropies, 
 respectively.
%
%
%
%
 The peak luminosity decreases as the fraction of C0 component within
 the cone becomes smaller.
 Hence,  
 although the values can vary by several factors,
 the observer views a brightest emission
 with a
 luminosity roughly equal to the predicted value
 when the LOS is located within the CO region.
 Accordingly, when the width
 of the C0 region $d\theta_0$ is larger,
 the probability for the observer to see the brightest region increases. 
 On the other hand,
 the luminosity drops significantly
 when the LOS is within the C1 region.
 As  seen in the figures,
 the difference in luminosity between the two range of observer
 is $\sim 2$ orders of magnitude.
%
%
%
%

%
 Regarding  spectra below the peak,
 softening from the case of the uniform jet ($\alpha_{\rm ph} \sim 0$) is seen
 due to the superposition of multiple components
(multi-color effect)
 as in the case of the two-component jet.
 Particularly, 
 this effect is pronounced
 when the LOS is located within the C1 region,
 since the contributions from
 C1 regions which enhance the emissions below the peak
 becomes larger.
 Regarding the dependence on the jet structure,
 the softening effect is more prominent 
 when the width of the C0 component $d\theta_0$ is 
 smaller, since the fraction C1 regions
 around the LOS tends to increase.
%
%
 In all cases,
 the observed low energy photon indices are roughly
 in the range $-1 \lesssim\alpha_{\rm ph}\lesssim 0$.\footnote{It is worth noting that, further softening can be 
 provided by the time evolution of the jet 
 as shown in Paper I.}


 The above results  show the
 spectral features,  
 the high and low energy slopes ($\beta_{\rm ph}$ and $\alpha_{\rm ph}$) 
 and peak (break) energy, that 
 resemble typical observations. 
 As mentioned before, 
 this implies that 
 typical observed spectra of GRB prompt emission
 can  indeed be reproduced regardless of the observer angle.
 It is stressed that multi-component structure  
 is favored
 for both the high energy and
 low energy spectra.
 To sum up,
 while the high energy spectra are produced by the accelerated 
 photons at the boundary layers,
 the low energy spectra are produced by the 
 superposition of thermal emissions from multiple components.
%
%
%
%
 We also emphasize that 
 the existence of the cut-off at $h\nu \sim \eta_0 m_e c^2 \sim 200 \eta_{0,400}~{\rm MeV}$
 is consistent with the recent observation by LAT/{\it Fermi}, 
 in the sense that these observations favor
 distinct emission components at 
 energies below and above $\sim 100~{\rm MeV}$.
%


 As in the case of the two-component jet,
 large asymmetry in the emission region
 produced by the boundary layer 
 is the main origin of the polarization signal in the multi-component jet.
 Reflecting the existence of the multiple boundary layers 
 (Fig.~\ref{modeleta}),
 the distribution of 
 the polarization signal $Q/I$ as a function of observer angle
 shows bumpy features
 (right panels of Figs.~\ref{m400-100}-\ref{m200-50}).
 The high energy non-thermal photons tend to 
 have higher polarization degree $|Q|/I$, since the asymmetry is larger.
 The general features of the polarization do not vary
 largely between 
 the cases of the high (Fig.~\ref{m400-100})
 and the low dimensionless entropies
  (Fig.~\ref{m200-50}),
 but the pronounced polarization degree  due to the non-thermal emissions  
 is found at lower energies in the latter case than in the former case
 since thermal peak energy is lower.

 The observer dependence of
 the overall polarization signal $Q/I$ can be
 roughly understood as follows :
 there is no polarization signal 
 for an on-axis observer $\theta_{\rm obs}=0$
 in any case, since the emission region is axisymmetric around the LOS.
 Regarding an off-axis observer  $\theta_{\rm obs}>0$,
 since
 the emissions originated
 in the C0 regions dominate over those in C1 regions
 at any observer angle,
 the distribution of C0 regions within 
 the cone of half-opening angle $\eta_0^{-1}\sim 0.14^{\circ}\eta_{0,400}^{-1}$
 around the LOS governs the polarization properties.
%
%
%
%
%
 When the LOS is within the C1 region,
 $Q/I$ is negative in all cases, since C0 regions within the cone
 is localized in the edge regions that produce negative $Q$.
 When the LOS is within the C0 region,
 $Q/I$ is always positive
 in the cases where $d\theta_0 \lesssim \eta_0^{-1}$ is satisfied
 (lower two panels of
 Figs.~\ref{m400-100} and \ref{m200-50}).
 This is due to the fact that the beaming cone 
 is not fully occupied by C0 region but
 localized in the region near the LOS
 that produces positive  $Q$ photon flux.
 On the other hand,
 in the cases of wider C0 regions
 $d\theta_0 \gtrsim \eta_0^{-1}$ 
(upper two panels of 
Figs.~\ref{m400-100} and \ref{m200-50}), 
 $Q/I$ is positive when 
 the LOS 
 is near the boundary due to the same reason, but
 becomes negative near the midpoint of C0 regions.
 This is  because the
 emissions near edge regions of  
 C0 component  brighter than the midpoint due to
 the photon acceleration effect.
 As a result, when the LOS is near the midpoint of
 C0 regions, photon flux within the beaming cone
 is dominated by the 
 contribution from edge regions that emit negative 
 $Q$ photons.

 As  seen in the figures,
 non-negligible degree of polarization $|Q|/I   \gtrsim {\rm few}~\%$
 is  present in large fraction of observer angle
 in all cases.
 These results suggest that
 significant polarization degree 
 is an inherent feature of the photospheric emissions
 from a jet that reproduce a typical observed spectra.
%
%
%
%
%
%
 The polarization degree is most pronounced at
 high energies above the thermal peak
 ($h\nu \gtrsim 1~{\rm MeV}$) and 
 can be higher than  $\sim 30\%$.
%
%
 Although relatively weak ($<20\%$),
 it is emphasized that 
 significant polarization degree 
  $\gtrsim 10\%$ can also be found at
 the energy bin ($h\nu \sim 100~{\rm keV}-1~{\rm MeV}$).
 This is particularly important, since
 the energy bin
 is  relevant to 
 the recent and future polarimeters such as
 GAP ($70~{\rm keV}-300~{\rm keV}$),
 TSUBAME ($30~{\rm keV}-200~{\rm keV}$)
 and POLAR ($50~{\rm keV}-500~{\rm keV}$).

\section{DISCUSSIONS}
\label{discussions}

\subsection{On the Jet Structure}
\label{stdis}

In the previous section, we have shown that
a broad non-thermal spectra are produced
in a photospheric emission from a stratified jet.
Particularly, 
it is demonstrated that
a multi-component jet that 
has interfaces of strong velocity gradient 
closely spaced within a small angular scale 
$\lesssim 2\Gamma^{-1}$
can   reproduce the typical Band spectra irrespective to the 
observer angle.
Although, we only explored the cases of
relatively narrow jet ($\theta_{\rm j}=1^{\circ}$ and $2^{\circ}$),
same results are expected
 for a wider jet, if similar structure 
develops in the entire jet.

The multi-component structure considered in the present study is
motivated by the
recent  hydrodynamical simulations
 of relativistic jet which 
show that the velocity shear regions naturally 
develop within the transverse structure of the entire jet
  during its propagation
\citep[e.g.,][]{MM13a, MM13b}.
Although the structure is  more complex,
we expect that the
properties of the resulting emissions
such as broad non-thermal spectra and high polarization 
degree  
are similar to the multi-component jet model. 
On the other hand, 
it is not clear whether the 
structure of the jet is 
naturally regulated to such a geometry which can reproduce the 
typical Band function.
This is particularly important for the 
high energy spectra,
since the photon acceleration is quite sensitive
to the velocity gradient as shown in \S\ref{Two}.
In order to fully explore this issue,
structure of the jet must be resolved in
an angular scale smaller than $\sim \Gamma^{-1}$.
Therefore,
simulations that follow the evolution of 
the jets in three-dimensions with
extremely high spacial resolution are required.
This is beyond the scope of the present study.

\subsection{Comparison with Previous Studies}
\label{prestudy}

In the present study, 
we have shown that the a significant polarization
is an inherent feature of photospheric emissions from a stratified jets.
Recently, \citet{LPR14} have performed a similar study
on the photospheric emissions.
The setup of their calculation is basically the same as ours, but
the imposed structure in the jet is different.
While we considered a multi-component jets having 
sharp velocity gradient in its interfaces,
they considered a 
smoothly decaying velocity profile 
in lateral direction
at the outer regions of the jet  ($\Gamma \propto \theta^{-p}$).
As in the present study,
they also found that a
broad non-thermal spectrum and
 significant polarization signal
can be realized 
in the resulting  emissions.
It is noted, however, in their model,
 that high polarization degrees 
$\gtrsim 10\%$ can be 
detected only by an observer that has LOS
aligned in the outer regions.
The corresponding observed luminosities 
are much dimmer than that observed by an
on-axis observer which views 
brightest emissions by an orders of magnitude.
%
Moreover, the shapes of the spectra 
have strong dependence on the observer angles,
and can deviate largely from the 
typical observations
in a wide range of observer angles 
particularly at high energies.
%
The main difference from their study
and the most important findings of the present study are that
the photospheric emissions associated in 
the stratified jet model
 can reproduce the typical observed spectra
irrespective to the observer angles,
and a high polarization degree
$\gtrsim 10\%$
can be found
not only in the observers that view dim emissions, 
but also those viewing brightest emissions.



%
Let us also consider the comparison with 
the synchrotron emission.
The polarization properties of the synchrotron emission 
have been extensively studied previously \citep{LPB03, G03,  NPW03, W03,  TSZ09, L06, ZY11}.
In the case of the synchrotron emissions,
 configuration of the magnetic field as well as 
the structure of the jet determine
the observed polarization signal.
Globally ordered  magnetic field configuration
(e.g., helical magnetic fields around the jet axis)
and wide opening angle of jet ($\theta_{\rm j}\gg \Gamma^{-1}$) are
favored to produce
large polarization degree \citep{LPB03, G03, TSZ09}.
Roughly speaking, the resulting polarization degree 
can be as high as $\gtrsim 40{\%}$ 
at large fraction of observer angle including
those viewing the brightest emissions \citep{Toma13}.
On the other hand, 
although high polarization degree up to $\sim 40\%$ can
be achieved at higher frequencies,
our results suggest that 
the polarization degree does not likely exceed
$\sim 20{\%}$ at $h\nu \lesssim 1~{\rm MeV}$ 
(energy ranges that are relevant for polarimetry observations)
in the photospheric emissions.
Hence,
if  the  detection polarization signal
is confirmed at higher level ($>20\%$),
synchrotron emission is favored for the emission mechanism of GRBs.
%
It should be noted, however, that the
 high  polarization degree found in the synchrotron emission model
is a result of the idealized assumptions such 
as ordered magnetic field and uniform jet structure.
Disruption of magnetic field configuration and/or
jet structures tend to reduce the net polarizations significantly.
%

%
%
%
%

%



\subsection{Comparison with Observation of Polarization}

As discussed in \S\ref{prestudy},
observations of polarization may give 
crucial constraint to the emission mechanisms of GRBs.
%
Up to now, 
there are only few reports for the detection of polarization
but most of the results are considered to be controversial,
since instrumental systematics cannot be ruled out 
\citep{CB03, RF04, WHA04, KBK07, MCD07, GLL09, MFM09, GCF13}.
Among them,
the most reliable measurement is provided by the 
recent observations by GAP instrument on board  IKAROS. 
In the observation, they detected
quite high degree of linear polarization 
in the three bright GRBs
which are
 GRB 100826A ($27\pm 11 \%$),
GRB 110301A ($70 \pm 22 \%$), and GRB 110721A ($84^{+16}_{-28} \%$)
 \citep{YMG11, YMG12}.
If such  high polarizations
($> 20{\%}$) are confirmed at the high confidence level,
 synchrotron emission
(or other optically thin emission models)
may be preferred for the emission mechanism \citep{Toma13}
rather than the photospheric emissions as discussed above. 
It is noted, however,
that the
measurements have large uncertainty due to the lack of photon statistics
and are still consistent
with unpolarized photons at $\sim 3\sigma$  confidence level. 
Therefore,  
robust discussion using the polarization measurement
is not possible at present.
Future missions such as TSUBAME \citep{YHK12} and POLAR \citep{OPOL11}
may help to resolve these issues.

In addition to the indication of high degree of polarizations,
GAP also reports on the time evolution in  
the polarization angle
(the direction of electric vector of the polarized emission) 
in one of the observed bursts.
%
When the observed data are split into two time
intervals, a shift in 
the polarization angle
 of $\sim 90^{\circ}$ was found for  GRB 100826A.
%
It is worth noting  that this rotation
can be explained 
within the framework of our model.
%
As shown in \S\ref{Multi},
the polarization signal $Q/I$ has a bumpy 
dependence on the observer angle and
changes signs ($\pm$) rapidly within a 
 small angular scale $\sim \Gamma^{-1}$.
Since the positive and negative $Q$
photons flux have
polarization angle perpendicular to each other,
the shift in $90^{\circ}$ can be result from 
the time variability in the jet properties
such as change in the width of each component
and/or bends in the jet
which effectively changes the observer angle
\citep[similar discussion  is also given in][]{LPR14}.

\section{SUMMARY AND CONCLUSIONS}
\label{summary}

In the present study, we have explored  spectral and polarization 
properties of photospheric emission from 
ultra-relativistic jets which have a structure in the transverse direction.
For the jet structure, we considered 
two-component and multi-component outflows
that have sharp velocity shear regions between each component.
The fluid properties such as electron number density $n_{\rm e}(r)$
and bulk Lorentz factor $\Gamma(r)$ 
are determined by applying the
adiabatic fireball model 
in each region independently.
Initially, unpolarized thermal photons are injected at a radius far below the
photosphere ($\tau(r_{\rm inj} \gg 1)$).
Using a Monte-Carlo technique,
we solve the evolution of the energy and polarization state of the 
injected photons via electron scatterings
until they reach the outer boundary located
far above the photosphere ($\tau(r_{\rm out}) \ll 1$).
 Below we summarize the main results and conclusions of the present study:

\noindent{1.} 
While the majority of the injected photons
escapes from the photosphere as 
adiabatically cooled thermal photons, 
small fraction of photons gains energy 
via Fermi-like acceleration mechanism
by crossing the velocity shear regions multiple times.
As a result, the accelerated photons produce a
non-thermal tail above the thermal peak energy in the 
observed spectra.
The non-thermal tail becomes harder as
the velocity gradient in the shear region
becomes larger
due to the increase in the acceleration efficiency.
Regarding the observer dependence, the 
non-thermal tail is most prominent when the 
line of sight of the observer 
is aligned to the velocity shear region.

\noindent{2.}
The presence of stratified structure within the jet 
produces large asymmetry in the emission region
around the line of sight of the observer.
As a result, polarization signal is inevitably accompanied
in the photospheric emission.
The polarization degree tends to increase as the velocity gradient
increases, since the asymmetry in the emission region is enhanced.
Regarding the energy dependence, 
emissions at high energies tend to show higher polarization degree
than those at lower energies.
This is because the emissivity of high energy (non-thermal) photons 
has stronger dependence on the lateral position ($\theta$) 
than that of the lower energy photons, 
since the high energy photons
produced by the photon acceleration process are
concentrated  
near the narrow velocity shear regions.


\noindent{3.}
Regardless of the observer angle,
photospheric emission from a multi-component  jet
can reproduce the typical observed spectra when the 
velocity shear regions are
spaced within an angular scale $\sim 2\Gamma^{-1}$.
%
Prominent non-thermal tail that has photon index 
similar to the typical value $-2.5 \lesssim \beta_{\rm ph} \lesssim -2$
can be present at any observer angle,
 since the velocity shear regions 
 are always located within an angle
 $\sim \Gamma^{-1}$ from the LOS.
Meanwhile, 
the spectrum below the peak energy is also modified from
the pure thermal one, 
since any observer can view the thermal photons that are
originated in the different components in the jet which 
have different peak energies.
This multi-color effect leads to softening which can result in 
a photon indices similar to the observed ones
$-1 \lesssim \alpha_{\rm ph} \lesssim 0$.
The maximum energy of the accelerated photons are 
limited by Klein-Nishina effect, and therefore 
the spectrum shows a cut-off at $\sim 200(\Gamma / 400) m_e c^2$
This 
is also  consistent with the recent observation by LAT/{\it Fermi}, 
which favor
distinct emission components at 
energies below and above $\sim 100~{\rm MeV}$.

\noindent{4.}
The multi-component jet
that reproduces typical observed spectra
also accompanies a 
non-negligible polarization signal ($|Q|/I   \gtrsim {\rm few}~\%$)
in a large fraction of observer angle. 
 The polarization degree is most pronounced at
 high energies above the  peak energy ($h\nu \gtrsim 1~{\rm MeV}$) 
 and 
 can be  higher than $\sim 30\%$.
%
%
 Although relatively weak ($<20\%$), 
 significant polarization degree 
  $\gtrsim 10\%$ can also be found at
 the energy bins ($h\nu \sim 100-1~{\rm MeV}$)
 which are relevant to 
 the recent and future polarimeters such as
 GAP ($70~{\rm keV}-300~{\rm keV}$),
 TSUBAME ($30~{\rm keV}-200~{\rm keV}$)
 and POLAR ($50~{\rm keV}-500~{\rm keV}$).


\acknowledgments 

This work is supported by
the
Japan Society for the Promotion of Science (No. 23340069 and
No. 25610056).
J.M. acknowledges support from
Grants-in-Aid for Foreign JSPS Fellow (Number 24.02022).
M.G.D acknowledges support from
Grants-in-Aid for Foreign JSPS Fellow (Number 25.03786).
S.L. acknowledges support from
Grants-in-Aid for Foreign JSPS Fellow (Number 25.03018).
We are grateful to Akihiro Suzuki for
fruitful discussions.
Numerical and data analysis were in part carried out on PC cluster at Center for Computational Astrophysics, National Astronomical Observatory of Japan.

\begin{appendix}
\section{METHOD FOR THE CALCULATION OF POLARIZATION}
\label{MCPOL}
Here we describe how the
polarization effect is taken into account in the scattering process
self-consistently in our calculation.
It is noted that
the method is basically identical to that used in \citet{LPR14}.
In our code,  each photon packet carries four-momentum $P^{\mu}= (h\nu / c, h\nu /c {\bf n})$ (or equivalently, the frequency $\nu$ and the propagation direction ${\bf n}$)
and Stokes parameters
\begin{eqnarray}
S= 
\begin{pmatrix}
I \\
Q \\
U \\
V
\end{pmatrix} .
\end{eqnarray}
Here, the parameter $I$ is set to be equal to the
total energy carried by the corresponding photon packet. 
The remaining parameters are always defined in a coordinate system
in which the
 $z-$axis is parallel to the photon propagation direction ${\bf n}$.
We use the convention that $Q/I=1$ ($Q/I=-1$) corresponds to a $100\%$ 
linear polarization parallel to the $x-$axis ($y-$axis) of the
coordinate system, $U/I=1$ ($U/I=-1$) corresponds to a $100\%$ 
linear polarization in the direction pointing at $45^{\circ}$ from
the $x-$axis in the anti-clockwise (clockwise) direction
 and 
$V/I=1$ ($V/I=-1$) corresponds to
 a full right (left) handed circular polarization.\footnote{It is noted that the parameters $Q$ and $V$  have opposite signs
 compared to those of \citet{LPR14}
 merely due to differences in their definition.
 Due these this differences,
 there are also slight changes in the
 formulas of the
 differential cross section for
 the scatterings
 and the scattering matrix 
 shown in Eq.~(\ref{dsdO}) and  (\ref{scmatrix}), respectively,
 from those used in their study.}
 As mentioned in \S\ref{MCcode},  while the global propagation of photons 
 is computed in the laboratory frame, 
 the scattering process which  changes
 the four momentum and polarization state are calculated in the
 fluid (electron) comoving frame.
 Hence, in each scattering event, we first perform Lorentz transformation
 of the photon four-momentum and Stokes parameters
 from the laboratory frame to the comoving frame. 
 Then, based on  the obtained quantities in the comoving frame,
 the four-momentum and  Stokes parameters of the scattered photons 
 are calculated \citep[for details of the Lorentz transformation, 
 see, e.g.,][and references therein]{LPR14}.
 Hereafter, we focus on the description of the calculations 
 performed in the comoving frame, and all
 physical quantities are evaluated in the corresponding frame.

\begin{figure}[ht]
\begin{center} 
\includegraphics[width=7cm]{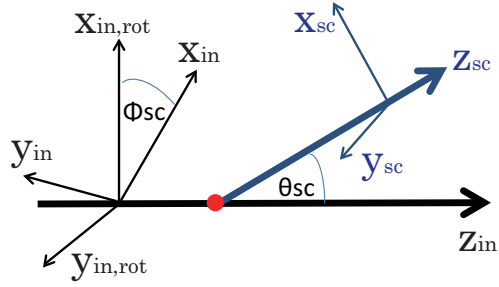}
\caption 
{The coordinate systems that are  used to calculate the scattering process
 in the fluid (electron) comoving frame.
The $z_{\rm in}$- and $z_{\rm sc}$-axes are
 the propagation direction of the incident and scattered photon, respectively,
and $\theta_{\rm sc}$ is the angle between the two axes.
The three $x-y$ coordinate systems are introduced to define the 
Stokes parameters of the incident and scattered photon. 
The initial Stokes parameters of the incident photon
$S_{\rm in}$ is determined in the 
 $x_{\rm in}$-$y_{\rm in}$ coordinate system.
The azimuthal angle $\phi_{\rm sc}$ is the
 angle between the scattering plane ($z_{\rm in}$-$z_{\rm sc}$ plane) 
and the $x_{\rm in}$-axis.
The $x_{\rm in, rot}$- and $y_{\rm in, rot}$-axes are defined as 
the coordinate system  obtained by rotating the $x_{\rm in}$- and
$y_{\rm in}$-axes around the $z_{\rm in}$-axis in the anti-clockwise direction
by an angle $\phi_{\rm sc}$.
Hence, the $x_{\rm in, rot}$-$z_{\rm in}$ plane is aligned to
the plane of scattering, and 
the rotated Stokes parameters of the incident photon $S_{\rm in, rot}$
is determined in this coordinate system.
The coordinate system used to define the Stokes parameters of the 
scattered photon $S_{\rm sc}$ is shown by 
the $x_{\rm sc}$- and $y_{\rm sc}$-axes.
The directions of the $x_{\rm sc}$- and $y_{\rm sc}$-axes are
determined in such a way that the $x_{\rm sc}$-$z_{\rm sc}$ plane is aligned
to the plane of scattering.
} 
\label{scfig}
\end{center}
\end{figure}

In Fig.~\ref{scfig}, we show the coordinate systems that are used to define
the Stokes parameters of the incident and scattered photons.
In each scattering event, the propagation direction of the photon 
after the scattering is determined based on the 
differential cross section of the electron scattering
that depends
 on the polarization state of the incident photon
\citep[e.g.,][]{BR78}.
While the Klein-Nishina effect is taken into account for 
an incident photon that has energy $h\nu_{\rm in}$ above $100~{\rm MeV}$, 
we neglect the effect in lower energies.
Hence, 
for a given set of Stokes parameters
(defined in the $x_{\rm in}$-$y_{\rm in}$
 coordinate system shown in Fig.~\ref{scfig}), $S_{\rm in}$,
we employ the differential cross section given by
\begin{eqnarray} 
\frac{d\sigma}{d\Omega}(\theta_{\rm sc}, \phi_{\rm sc})=\frac{r_0^2}{2}
\left(\frac{\nu_{\rm sc}}{\nu_{\rm in}}\right)^2
\left[ \left(\frac{\nu_{\rm in}}{\nu_{\rm sc}}\right)^2 + \left(\frac{\nu_{\rm sc}}{\nu_{\rm in}}\right)^2
 - \sin^2{\theta_{\rm sc}} 
 \left\{ 1+ \left(\frac{Q}{I} \right)\cos{2\phi_{\rm sc}} + \left(\frac{U}{I} \right) \sin{2\phi_{\rm sc}}
 \right\}
 \right],
 \label{dsdO}
\end{eqnarray}
where  $r_0$ is the classical electron radius, and
 $\nu_{\rm sc}$ is the frequency of the scattered photon determined as
\begin{eqnarray}
\label{nusc}
\nu_{\rm sc} = \left\{ \begin{array}{ll}
 \nu_{\rm in} &~~
         {\rm for}~~ h\nu_{\rm in} < 100~{\rm MeV}  ,  \\
     \nu_{\rm in}/[1 + \nu_{\rm in}(1- \cos{\theta_{\rm sc}})]  &~~
         {\rm for}~~  h\nu_{\rm in}\geq 100~{\rm MeV}  . \\
             \end{array} \right. 
\end{eqnarray}
Here, $\theta_{\rm sc}$ and  $\phi_{\rm sc}$ are the 
 angles between the propagation directions
of the photon before and after scattering and 
 between the
scattering plane and the $x$-axis
 ($x_{\rm in}-$axis shown in Fig.~\ref{scfig}) of
 the coordinate system used to 
define the Stokes parameters of the incident photon, respectively.
In the above equation,
we assumed averaging over the isotropic electron spin distribution.

Once the scattering angles ($\theta_{\rm sc}$ and $\phi_{\rm sc}$)
 are determined,
we update the four-momentum of the photon by
replacing it with that of the scattered photons.
Then, we update the Stokes parameter by
calculating the Stokes parameter of the scattered photon.
In this procedure,
it is convenient to 
 employ  coordinate systems that have their $x-z$ planes
 coinciding with the scattering plane to define the Stokes parameters
 for both the incident and scattered photons 
($x_{\rm in, rot}-$$y_{\rm in, rot}$
 and $x_{\rm sc}-$$y_{\rm sc}$ coordinate systems
 shown in  Fig.~\ref{scfig} for the incident and scattered photon,
 respectively).
Hence, we first
determine the Stokes parameters of the incident photon
defined in the new coordinate system
 ($x_{\rm in, rot}$-$y_{\rm in, rot}$ coordinate), $S_{\rm in, rot}$,
in which 
$x$-axis is pointed at an angle $\phi_{sc}$
in the anti-clockwise direction with respect to the
original one ($x_{\rm in}$-$y_{\rm in}$ coordinate).
The rotational transformation of the Stokes parameters
is performed as
$S_{\rm in, rot}= L(\phi_{\rm sc}) S_{\rm in}$, where 
$L(\phi_{\rm sc})$ is the 
rotation matrix given by
\citep{M61}
\begin{eqnarray}
L(\phi) = 
\begin{pmatrix}
1 & 0 & 0 & 0 \\
0 & \cos{2\phi} & \sin{2\phi} & 0 \\
0 & - \sin{2\phi} & \cos{2\phi} & 0 \\
0 & 0 & 0 & 1 
\end{pmatrix}.
\end{eqnarray}
Using the transformed Stokes parameters,
Stokes parameters after the scattering
defined in the  coordinate system that has
$x-z$ plane coinciding with the scattering plane, $S_{\rm sc}$,
can be calculated by using the 
scattering matrix \citep{M61}
\begin{eqnarray}
R(\theta_{\rm sc}) = 
{\scriptsize
\begin{pmatrix}
1+ \cos^2{\theta_{\rm sc}} + (\nu_{\rm in} - \nu_{\rm sc})(1- \cos{\theta_{\rm sc}}) & -  \sin^2{\theta_{\rm sc}} & 0 & 0\\
 - \sin^2{\theta_{\rm sc}}  & 1 + \cos^2{\theta_{\rm sc}} & 0 & 0  \\
0 & 0 & 2 \cos{\theta_{\rm sc}} & 0 \\
0 & 0 & 0 &  2 \cos{\theta_{\rm sc}} + (\nu_{\rm in} - \nu_{\rm sc})(1- \cos{\theta_{\rm sc}}) \cos{\theta_{\rm sc}}
\end{pmatrix},
}
\label{scmatrix}
\end{eqnarray}
as
$S_{\rm sc} = R(\theta_{\rm sc})S_{\rm in, rot}$.
It is obvious from the above matrix that a photon with $V=0$
does not obtain any circular polarization from the scattering as mentioned in 
\S\ref{MCcode}.
To sum up, for a given
set of
 scattering angles 
and initial Stokes parameters, 
the Stokes parameters of the scattered photon are
calculated as
\begin{eqnarray}
S_{\rm sc} = R(\theta_{\rm sc})L(\phi_{\rm sc})S_{\rm in}.
\end{eqnarray}
Finally, we normalize the obtained Stokes parameters 
so that the first component $I$ is  equal to the total energy of the 
photon packet.
Since the packet is treated  as an ensemble of photons having equal
frequency,
the change in the total energy  due to scattering 
is computed in the same way as its frequency
(Eq.~(\ref{nusc})).
Therefore,
while
the total energy of the scattered packet
decreases from that of the incident packet by a
factor $[1 + \nu_{\rm in}(1-\cos{\theta_{\rm sc}})]^{-1}$ 
for  $\nu_{\rm in} \geq 100~{\rm MeV}$, 
it remains constant for $\nu_{\rm in} < 100~{\rm MeV}$.

\end{appendix}

\end{document}